\documentclass[a4paper,fleqn,usenatbib]{mnras}

\usepackage{newtxtext,newtxmath}

\usepackage[T1]{fontenc}
\usepackage{ae,aecompl}

\usepackage{graphicx}	
\usepackage{amsmath}	
\usepackage{amssymb}	
\usepackage{breakurl}

\newcommand{\dg}{^{\circ}}

\title[RoboPol: EVPA rotations in blazars]{RoboPol: Connection between optical polarization plane
rotations and gamma-ray flares in blazars}

\author[D. Blinov et al.]
{D. Blinov$^{1,2,3}$\thanks{E-mail: blinov@physics.uoc.gr}, V. Pavlidou$^{1,2}$, I. 
Papadakis$^{1,2}$, S. Kiehlmann$^{4}$, I. Liodakis$^{5,1}$,
\newauthor
G. V. Panopoulou$^{6}$, E. Angelakis$^{7}$, M. Balokovi\'{c}$^{6}$, T. Hovatta$^{8}$, O. G. 
King$^{6}$,
\newauthor
A. Kus$^{9}$, N. Kylafis$^{2,1}$, A. Mahabal$^{6}$, S. Maharana$^{10}$, I. Myserlis$^{7}$, E. 
Paleologou$^{1,2}$,
\newauthor
I. Papamastorakis$^{1,2}$, E. Pazderski$^{9}$, T. J. Pearson$^{6}$, A. Ramaprakash$^{10}$,
\newauthor
A.\,C.\,S. Readhead$^{6}$, P. Reig$^{2,1}$, K. Tassis$^{1,2}$, J. A. Zensus$^{7}$ \\
$^{1}$Department of Physics and Institute for Theoretical and Computational Physics (ITCP), 
University of Crete, 71003, Heraklion, Greece\\
$^{2}$Foundation for Research and Technology - Hellas, IESL, Voutes, 7110 Heraklion, Greece\\
$^{3}$Astronomical Institute, St. Petersburg State University, Universitetsky pr. 28, Petrodvoretz, 198504 St. Petersburg,
Russia \\
$^{4}$Owens Valley Radio Observatory, California Institute of Technology, Pasadena, CA 91125, USA\\
$^{5}$KIPAC, Stanford University, 452 Lomita Mall, Stanford, CA 94305, USA\\
$^{6}$Cahill Center for Astronomy and Astrophysics, California Institute of Technology, 1200 E 
California Blvd, MC 249-17,\\Pasadena CA, 91125, USA\\
$^{7}$Max-Planck-Institut f\"{u}r Radioastronomie, Auf dem H\"{u}gel 69, 53121 Bonn, Germany\\
$^{8}$Tuorla Observatory, Department of Physics and Astronomy, University of Turku, Finland\\
$^{9}$Toru\'{n} Centre for Astronomy, Nicolaus Copernicus University, Faculty of Physics, Astronomy 
and Informatics, Grudziadzka 5,\\ 87-100 Toru\'{n}, Poland\\
$^{10}$Inter-University Centre for Astronomy and Astrophysics, Post Bag 4, Ganeshkhind, Pune - 411 
007, India\\
}

\date{Accepted XXX. Received YYY; in original form ZZZ}

\pubyear{2017}

\begin{document}
\label{firstpage}
\pagerange{\pageref{firstpage}--\pageref{lastpage}}
\maketitle

\begin{abstract}
We use results of our 3 year polarimetric monitoring program to investigate the previously 
suggested connection between rotations of the polarization plane in the optical emission of blazars 
and their gamma-ray flares in the GeV band. The homogeneous set of 40 rotation events in 24 sources
detected by {\em RoboPol} is analysed together with the gamma-ray data provided by {\em Fermi}-LAT. 
We confirm that polarization plane rotations are indeed related to the closest gamma-ray flares in 
blazars and the time lags between these events are consistent with zero. Amplitudes of the rotations 
are anticorrelated with amplitudes of the gamma-ray flares. This is presumably caused by higher 
relativistic boosting (higher Doppler factors) in blazars that exhibit smaller amplitude 
polarization plane rotations. Moreover, the time scales of rotations and flares are marginally 
correlated.
\end{abstract}




\section{Introduction} \label{sec:introduction}

Blazars are a subclass of active galactic nuclei with relativistic jets oriented toward the Earth. 
Due to the close alignment of jets with the line of sight, their emission is strongly 
relativistically boosted and prevails in the overall emission \citep{Blandford1979}. The broadband 
spectral energy distribution (SED) of a blazar has two broad humps, peaking in the IR--X-ray bands 
and in the MeV--TeV band. The low energy part of the SED is produced by relativistic electrons in 
the jet emitting synchrotron radiation. The nature of the high energy component is uncertain. For 
instance, it is unknown what kind of particles in the jet is responsible for upscattering of 
photons to gamma-ray energies \citep{Bottcher2013} and where the gamma-ray emitting site is located 
in the jet \citep[e.g.,][]{Poutanen2010,Agudo2011}.

Optical emission of blazars is often significantly polarized owing to its synchrotron origin. 
Despite the fact that optical fractional polarization is correlated with the total flux in optical 
and gamma-rays for some blazars \citep{Itoh2016}, ordinarily both the electric vector position 
angle (EVPA) and fractional polarization behave erratically \citep{Uemura2010}. However, a number 
of events have been reported, where the EVPA performs continuous and gradual rotations, whose 
amplitudes are as high as hundreds of degrees. In some cases these EVPA rotations occur together 
with flares in multiple bands \citep[e.g.,][]{Marscher2010,Aleksic2014}. The physical processes 
behind the EVPA rotations and their connection to gamma-ray emission of blazars remain unclear. 
There are a number of models proposed for the interpretation of such events. They can be divided in 
two general classes: random walk and deterministic models. The former class explains polarization 
rotations as occasional periods of smooth variability in EVPA curves produced by stochastic 
variations of the polarization vector \citep{Jones1985,Marscher2014,Kiehlmann2016}. Deterministic 
models describe EVPA variability by relativistic aberration 
\citep{Abdo2010,Larionov2013,Aleksic2014}, change of the magnetic field structure \citep{Zhang2016}, 
or magnetic field reconnections \citep{Deng2016} or other deterministic processes 
\citep{Lyutikov2017}. It has been suggested that both types of EVPA rotations can occur even in a 
single blazar \citep{Blinov2015,Kiehlmann2016}.

In order to increase the number of detected EVPA rotations and improve our understanding of this 
phenomenon we started the {\em RoboPol} programme\footnote{\url{http://robopol.org}}. It has been 
designed for efficient detection of EVPA rotations in a statistically meticulously defined sample of 
blazars \citep{Pavlidou2014}. Together with monitoring data provided by the Large Area Telescope 
(LAT) onboard the {\em Fermi} gamma-ray space observatory \citep{Atwood2009}, it provides an 
unrivalled opportunity to investigate the potential connection between optical EVPA variability and 
gamma-ray activity.

{\em RoboPol} started observations at Skinakas observatory, Greece, in May 2013. The EVPA rotations 
detected during its first three years of operation were presented in \citet[hereafter Papers~I, II 
and III]{Blinov2015,Blinov2016a,Blinov2016b}. In Paper~I we analysed a set of EVPA rotation events 
detected during the first observing season and their connection to gamma-ray flaring activity in 
blazars. We found that it is unlikely that all EVPA rotations are produced by random variability of 
the polarization plane. Moreover, it is very unlikely that none of the EVPA rotation events is 
connected with accompanying flares in the gamma-ray band. In the current paper we verify results of 
Paper~I using the entire set of rotation events detected in three years. We search for the existence
of correlations between parameters of EVPA rotations and gamma-ray flares. Such correlations are
expected if the two classes of events are physically connected.

The values of the cosmological parameters adopted throughout this work are $H_0 = 67.8$ km s$^{-1}$ 
Mpc$^{-1}$, $\Omega_m = 0.308$ and $\Omega_\Lambda = 1 - \Omega_m$ \citep{Ade2016}. In all the 
statistical tests we use a limit $p = 0.05$ as the acceptance limit.

\section{Observations and data reduction} \label{sec:observations}
\subsection{Optical Observations}

The observation routines used for acquisition of the optical data have been described in 
\cite{King2014} and Paper~I. Nevertheless, for reader's convenience we briefly summarize our
observations here. Our polarimetric and photometric measurements were performed using {\em RoboPol}
photopolarimeter, which was specifically designed for the project. The polarimeter is installed at
the 1.3-m telescope of the Skinakas observatory\footnote{\url{http://skinakas.physics.uoc.gr/}}.
The data analysed in this paper were taken during the 2013 - 2015 observing seasons. The monitored 
sample included 62 gamma-ray--loud and 15 gamma-ray--quiet sources. It has been selected on the 
basis of stringent, objective and bias-free criteria \citep[see][for details]{Pavlidou2014}.
The EVPA rotation events analysed in this paper have been reported in Papers~I -- III.

All data were obtained with the R-band filter. The exposure length was adjusted according to the 
brightness of each target, which was estimated during the short pointing exposures, depending also 
on the sky conditions. The data were processed using the specialized pipeline described in detail 
by \cite{King2014}.

The mean value of $E(B - V)=0.11^{\rm m}$ in the fields of our sources, suggests that typical 
interstellar polarization is less than 1\% \citep{Serkowski1975}. The EVPA is commonly defined
with accuracy between $1\dg$ - $10\dg$, however it strongly depends on the fractional polarization
of the source and its brightness. The polarization degree is typicaly measured with accuracy better
than 1\%. A detailed description of the instrument model and error analysis is given in
\cite{King2014}.

In order to resolve the $180\dg$ ambiguity of the EVPA we followed a standard procedure \citep[see 
e.g.,][]{Kiehlmann2013}, which is based on the assumption that temporal variations of the EVPA are 
smooth and gradual, hence adopting minimal changes of the EVPA between consecutive measurements.
This procedure is described in more detail in Papers~I and II.

\subsection{Gamma-ray Observations} \label{subsec:gamma_obs}

For the acquisition of the gamma-ray data we closely followed the procedure described in 
\cite{Blinov2015}. We analysed the {\em Fermi} Large Area Telescope (LAT) data. The
{\em Fermi} gamma-ray space observatory observes the entire sky every at energies of 20 MeV -- 
300 GeV every 3 hours in normal mode \citep{Atwood2009}. We processed the data in the energy range 
$100\, {\rm MeV} \le E \le 100\, {\rm GeV}$ using the unbinned likelihood analysis of the standard 
{\em Fermi} analysis software package Science Tools v10r0p5 and the instrument response function 
$P8R2\_SOURCE\_V6$. Source class photons (evclass=128 and evtype=3) were selected within a $15\dg$ 
region of interest centred on a blazar. A cut on the satellite zenith angle ($< 90\dg$) was used to 
exclude the Earth limb background. The diffuse emission from the Galaxy was modelled using the 
spatial model $gll\_iem\_v06$. The extragalactic diffuse and residual instrumental backgrounds were
included in the fit as an isotropic spectral template $iso\_source\_v05$. The background 
models\footnote{\url{http://fermi.gsfc.nasa.gov/ssc/data/access/lat/4yr_catalog/gll_psc_v16.xml}} 
include all sources from the 3FGL \citep[third {\em Fermi}-LAT source catalogue,][]{Acero2015} 
within $15\dg$ of the blazar. Photon fluxes of sources beyond $10\dg$ from the blazar and spectral 
shapes of all targets were fixed to their values reported in 3FGL. The source is considered to be 
detected if the test statistic, TS, provided by the analysis exceeds 10, which corresponds to 
approximately a $3\sigma$ detection level \citep{Nolan2012}. The systematic uncertainties in the 
effective LAT area do not exceed 10 per cent in the energy range we use \citep{Ackermann2012}. This 
makes them insignificant with respect to the statistical errors that dominate over the short time 
scales analysed in this paper. Moreover, our analysis is based on relative flux variations. 
Therefore the systematic uncertainties were not taken into account.

Different time bins $t_{\rm int}$, from 2 to 25 d were used, depending on the flux density 
of the object. In order to make the analysis more robust we increased sampling of the photon flux 
curves overlapping adjacent time bins. The centres of the bins were separated by $t_{\rm int}/4$ 
interval from each other. This prevents losses of possible short-term events in the light curves and 
reduces the dependence of results on the particular position of the time bins. The oversampling 
introduces an autocorrelation in the photon flux curves, which, however, does not affect the results 
presented in this work \citep[see also][]{Blinov2015}.

\section{The connection between EVPA rotations and gamma-ray flares} 
\label{sec:rot_gamma}

In Paper~I we reported on 16 EVPA rotation events detected in 14 blazars (hereafter 
``rotators'') during the 2013 observing season and investigated their connection to gamma-ray 
activity. Later, in Papers~II and III, we analysed statistical properties of 24 more rotations 
detected in 2014 and 2015. In the following sections we extend the analysis of Paper~I to the 
entire set of 40 EVPA rotations detected in 24 blazars (see Table A1 in Paper III) and investigate 
whether these events are related to the gamma-ray flaring activity of the blazars.

\subsection{Time lags between rotations and gamma-ray flares} \label{subsec:obs_fits}

We identify gamma-ray flares according to a formal definition of a gamma-ray flare similar to the 
one proposed by \cite{Nalewajko2013}: ``a flare is a contiguous period of time, associated with a 
given photon flux peak, during which the photon flux exceeds 2/3 of the peak value, and this lower 
limit is attained exactly twice -- at the start and at the end of the flare''.

\begin{figure}
 \centering
 \includegraphics[width=0.4\textwidth]{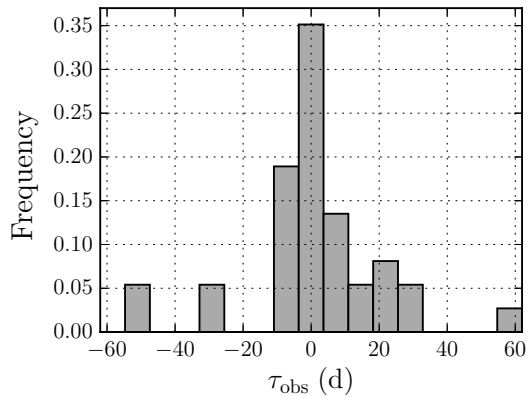}
 \caption{Distribution of observed time lags, $\tau_{\rm obs}$, between the {\em RoboPol} EVPA 
rotations and the Fermi gamma-ray flares.}
 \label{fig:tl_dist}
\end{figure}

We searched for gamma-ray flares within time intervals corresponding to {\em RoboPol} observing 
seasons of each blazar. The flares are marked by the red  points in the photon flux curves of 
rotators shown in Fig.~\ref{fig:rotations1} - \ref{fig:rotations3}. Then we fitted these events 
using profiles with an exponential rise and decay. This kind of profile is commonly used for fitting 
individual blazar flare pulses in optical, gamma and radio bands \citep[e.g.,][]{Abdo2010b}:
\begin{equation}\label{fit_func}
 F(t) = F_{\rm c} + \sum_{i=1}^{N} F_{\rm p,i} \left( \exp\left({\frac{t_{\rm p,i}-t}{T_{\rm 
r,i}}}\right) + \exp\left({\frac{t-t_{\rm p,i}}{T_{\rm d,i}}}\right) \right)^{-1}.
\end{equation}
$F_{\rm c}$ represents the constant photon flux level underlying the flares, $N$ is the number of 
flares,  $F_{\rm p,i}$ measures the amplitude of the $i$-th flare, $t_{\rm p,i}$ describes the time 
of the peak (it corresponds to the actual maximum only for symmetric flares), and $T_{\rm r,i}$ and 
$T_{\rm d,i}$ measure the rise and decay time, respectively. All the parameters were set to be free 
during the fitting procedure, while initial values used in the fitting procedure were estimated 
from the photon flux curves. The flares in RBPL\,J1927+6117 from the 2013 season and 
RBPL\,J1809+2041 and RBPL\,J1836+3136 from the 2015 season could not be fitted due to the poor 
sampling of the photon flux curves. In most of the cases all flares identified along the observing 
season were fitted together using equation \ref{fit_func}. However, in several cases (e.g., 2014 
curves for RBPL\,J1512-0905 and RBPL\,J1800+7828) a decent fit could not be achieved for all flares 
together or it resulted in physically meaningless values of parameters (e.g., negative fluxes). In 
these curves we fitted each gamma-ray flare separately. The best fits of the curves are shown by 
the blue lines in Fig.~\ref{fig:rotations1}-\ref{fig:rotations3}.

Time lags, $\tau_{\rm obs}$, between rotations and the closest gamma-ray flares were estimated as
$\tau_{\rm obs} = \overline{T^{\rm rot}} - t_{\rm p}$, where $\overline{T^{\rm rot}}$ is the middle 
point of each EVPA rotation, defined as the mean Julian Date (JD) of the first and the last points 
of the rotation. Fig.\ref{fig:tl_dist} shows the distribution of time lags.
The distribution has a clear peak around zero time lag. The mean of the distribution is 1.5 
d, while the standard deviation is 19 d. The Kolmogorov-Smirnov (K-S) test rejects the hypothesis 
that the distribution is consistent with the uniform distribution at the $0.4$\% confidence level. 
At the same time the distribution is consistent with the normal distribution ($p\text{-value} = 
0.03$) according to normality test by \cite{DAgostino1973}.

\begin{figure}
 \centering
 \includegraphics[width=0.46\textwidth]{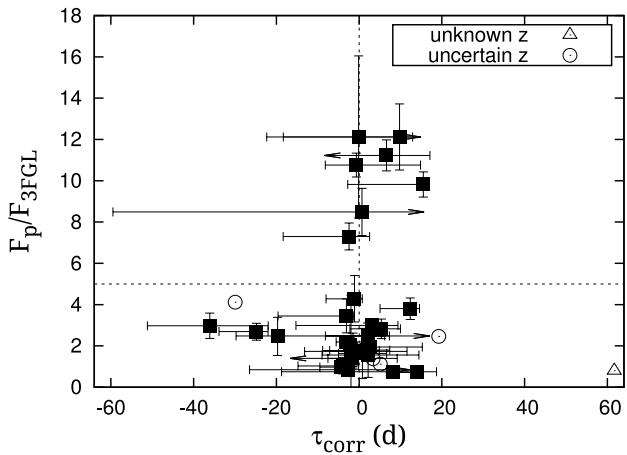}
 \caption{Normalized gamma-ray flare amplitude, $F_{\rm p}/F_{\rm 3FGL}$, versus time lags, 
$\tau_{\rm corr}$. Filled squares, open circles and triangles correspond to well defined, uncertain 
and unknown redshift. For the last category we use $\tau_{\rm obs}$. The horizontal dashed line 
indicates $F_{\rm p}/F_{\rm 3FGL}=5$, where we put the border between high and low amplitude 
flares.}
 \label{fig:TLvsAmplobs}
\end{figure}

\subsection{Observed gamma-ray flare amplitudes vs. timelags} \label{subsec:obs_ampl_tl}

In Paper~I we found that high amplitude gamma-ray flares are concurrent in time with EVPA 
rotations, while some low amplitude flares have time lags that are significantly different from 
zero. We investigate the existence of this trend with the full set of detected EVPA rotations. 
Following the analysis of Paper~I, we normalized the amplitude, $F_{\rm p}$, of the gamma-ray flare 
closest to the EVPA rotation event by the average photon flux, $F_{\rm 3FGL}$, of each blazar as 
listed in 3FGL. The relative amplitudes versus the time lags are plotted in 
Fig.~\ref{fig:TLvsAmplobs}. The filled black squares show redshift-corrected time lags i.e., 
$\tau_{\rm corr} = \tau_{\rm obs}/(1 + z)$, while open symbols represent blazars with either 
unknown or uncertain $z$ (see Paper~III for the list of redshifts). The distribution of points in 
Fig.~\ref{fig:TLvsAmplobs} shows possible bimodality along the amplitude axis, which suggests 
existence of two classes of flares: high- and low-amplitude. Following Paper~I, we arbitrary set the 
limit between the two classes at $F_{\rm 3FGL}/F_{\rm p} = 5$. The apparent tendency for the 
difference of time lags between these high and low amplitude flares, reported in Paper~I, remained 
for the full set of rotations. All eight brightest flares in Fig.~\ref{fig:TLvsAmplobs}, located 
above the dashed line, are consistent with $\tau_{\rm corr} = 0$. At the same time there are a 
number of low amplitude flares which appear to have non-zero time lags with respect to EVPA 
rotations.

\begin{figure*}
 \centering
 \includegraphics[width=0.494\textwidth]{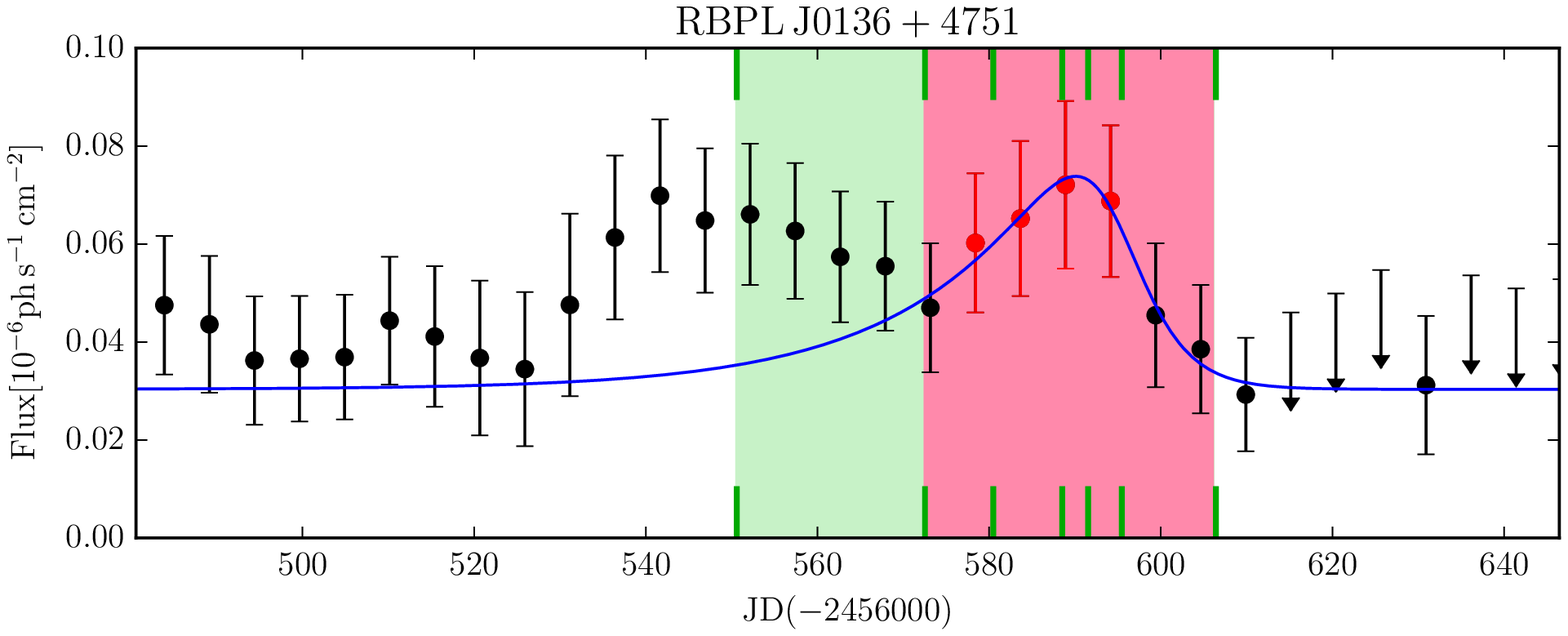}
 \includegraphics[width=0.494\textwidth]{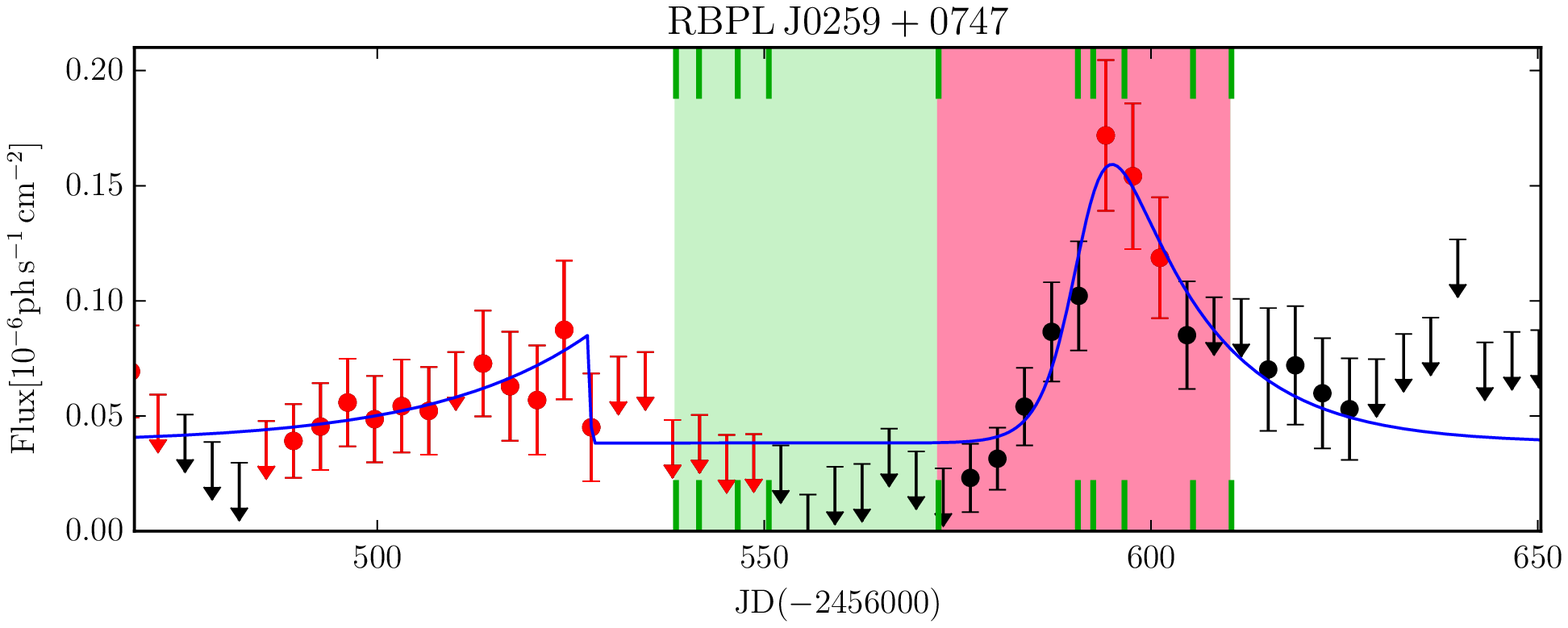}\\
 \includegraphics[width=0.494\textwidth]{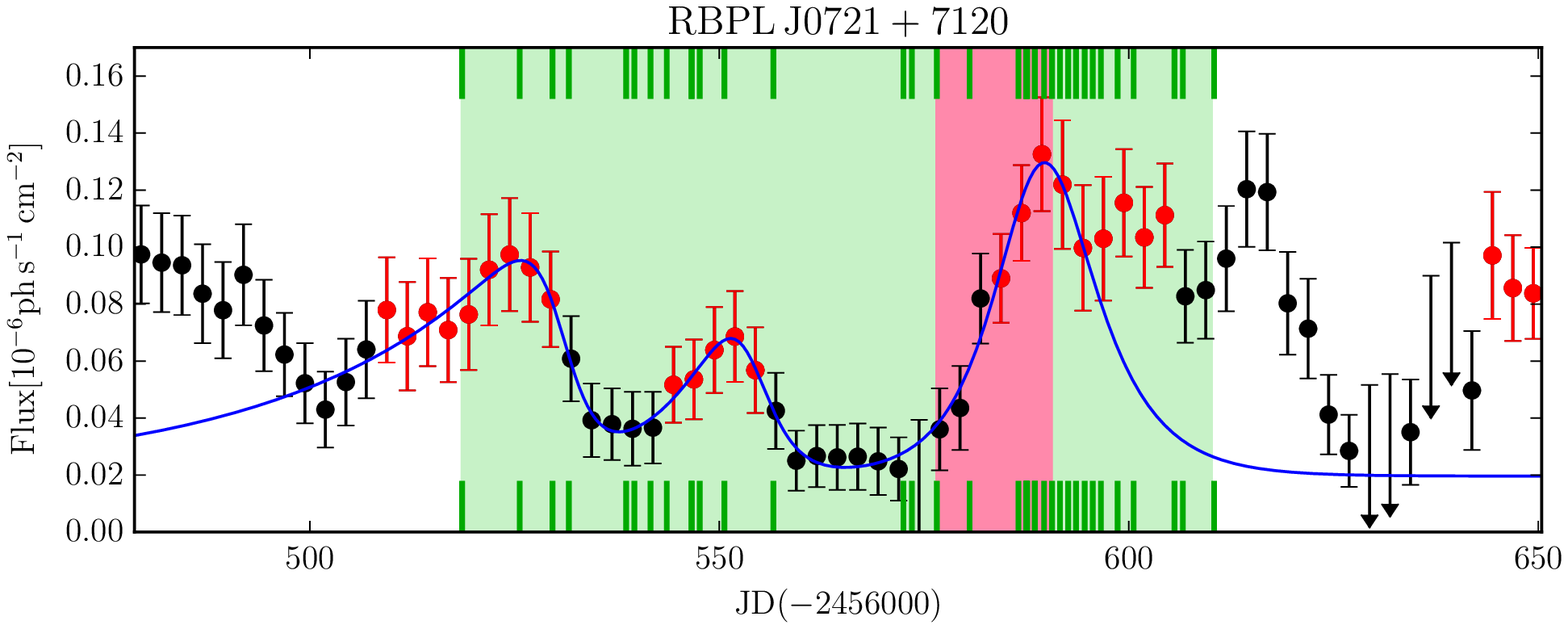}
 \includegraphics[width=0.494\textwidth]{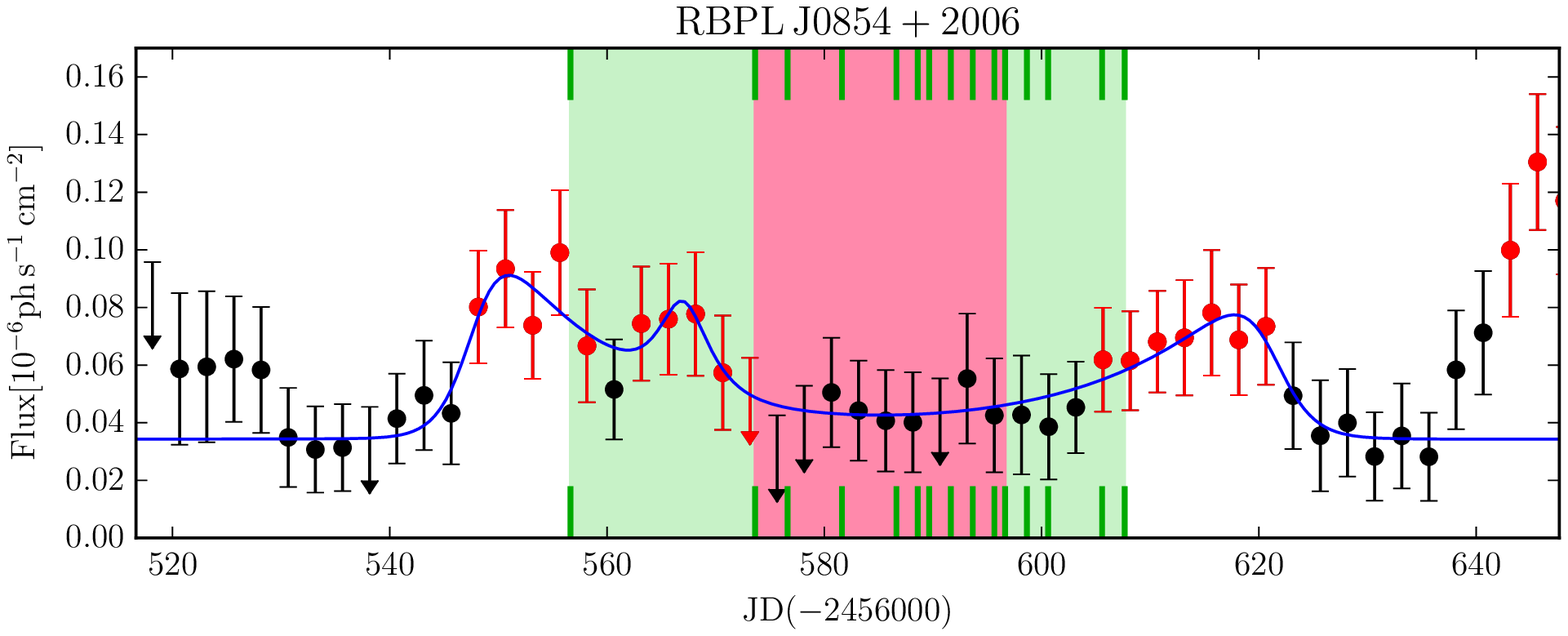}\\
 \includegraphics[width=0.494\textwidth]{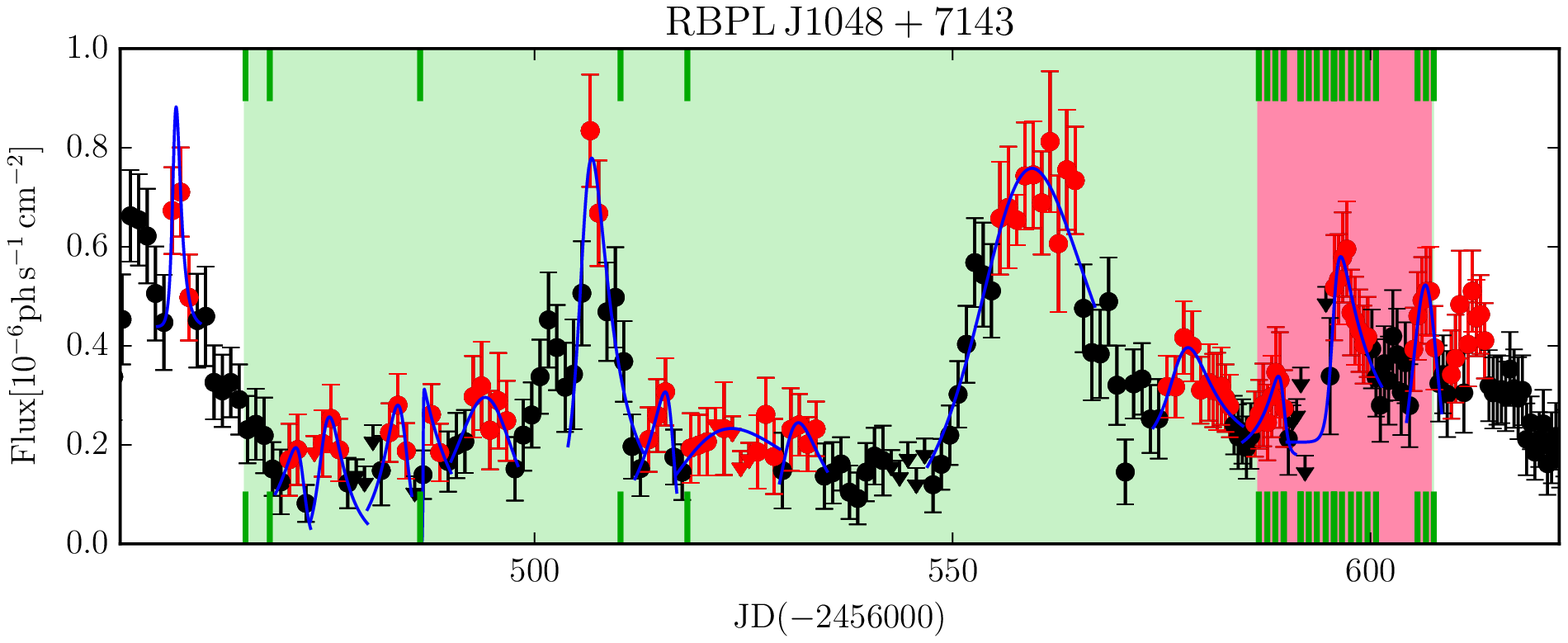}
 \includegraphics[width=0.494\textwidth]{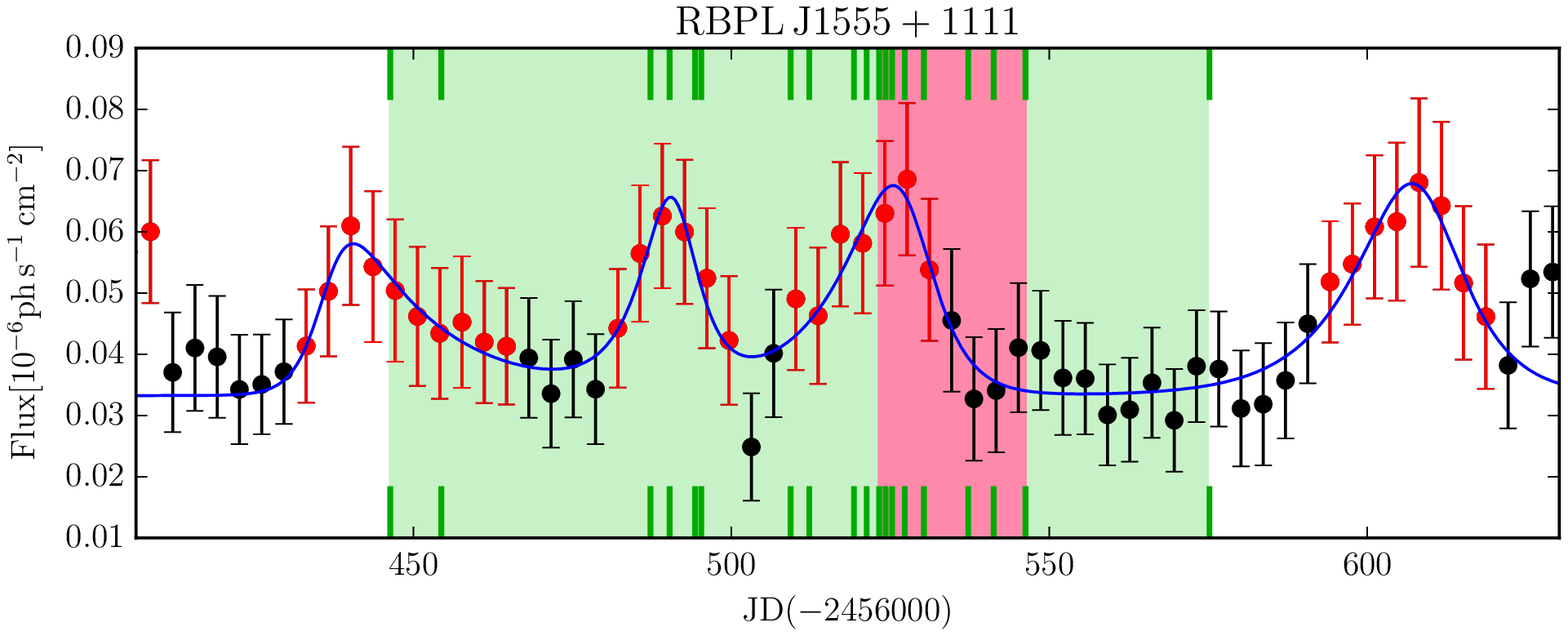}\\
 \includegraphics[width=0.494\textwidth]{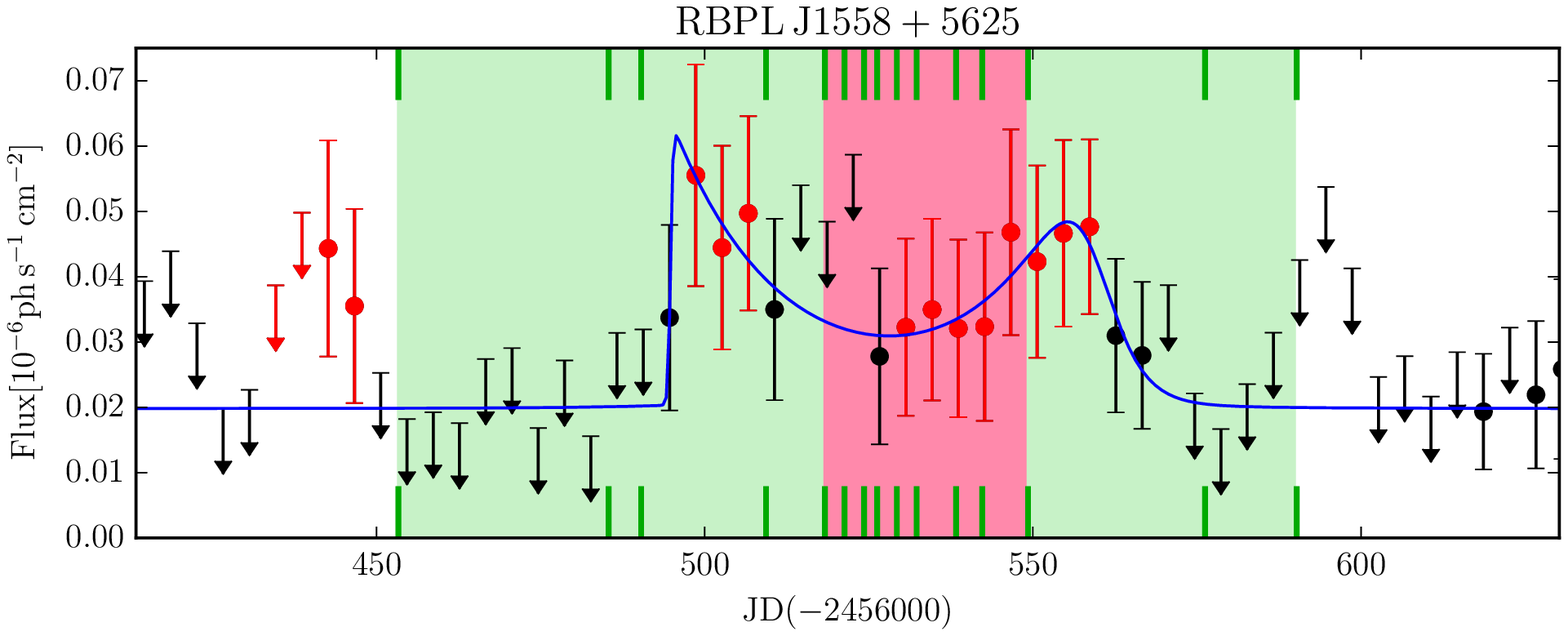}
 \includegraphics[width=0.494\textwidth]{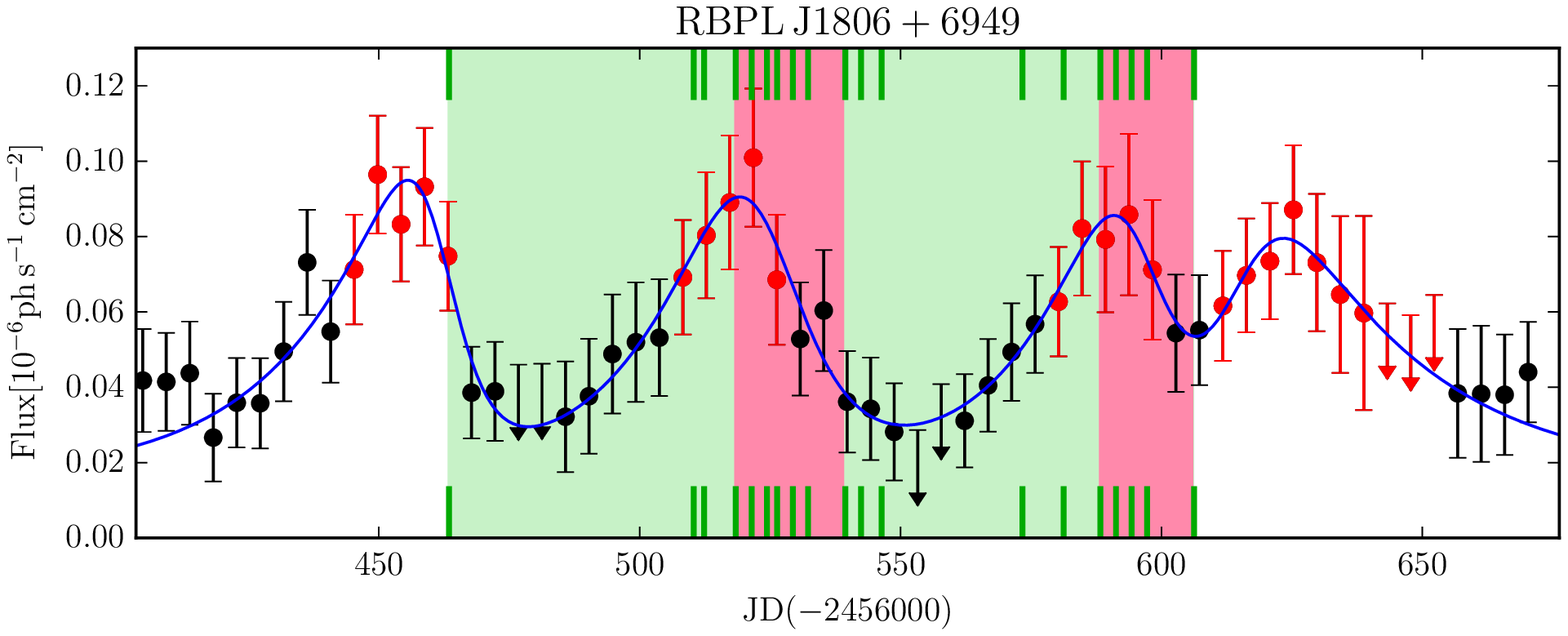}\\
 \includegraphics[width=0.494\textwidth]{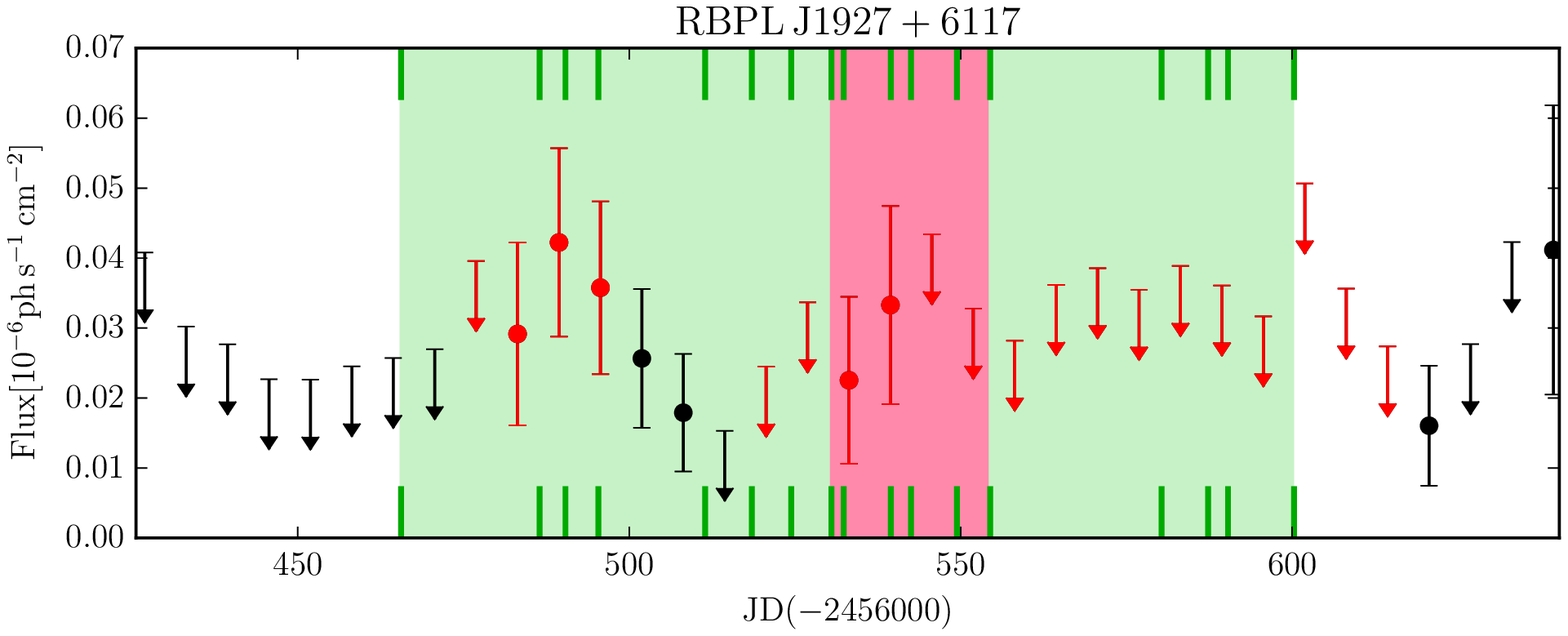}
 \includegraphics[width=0.494\textwidth]{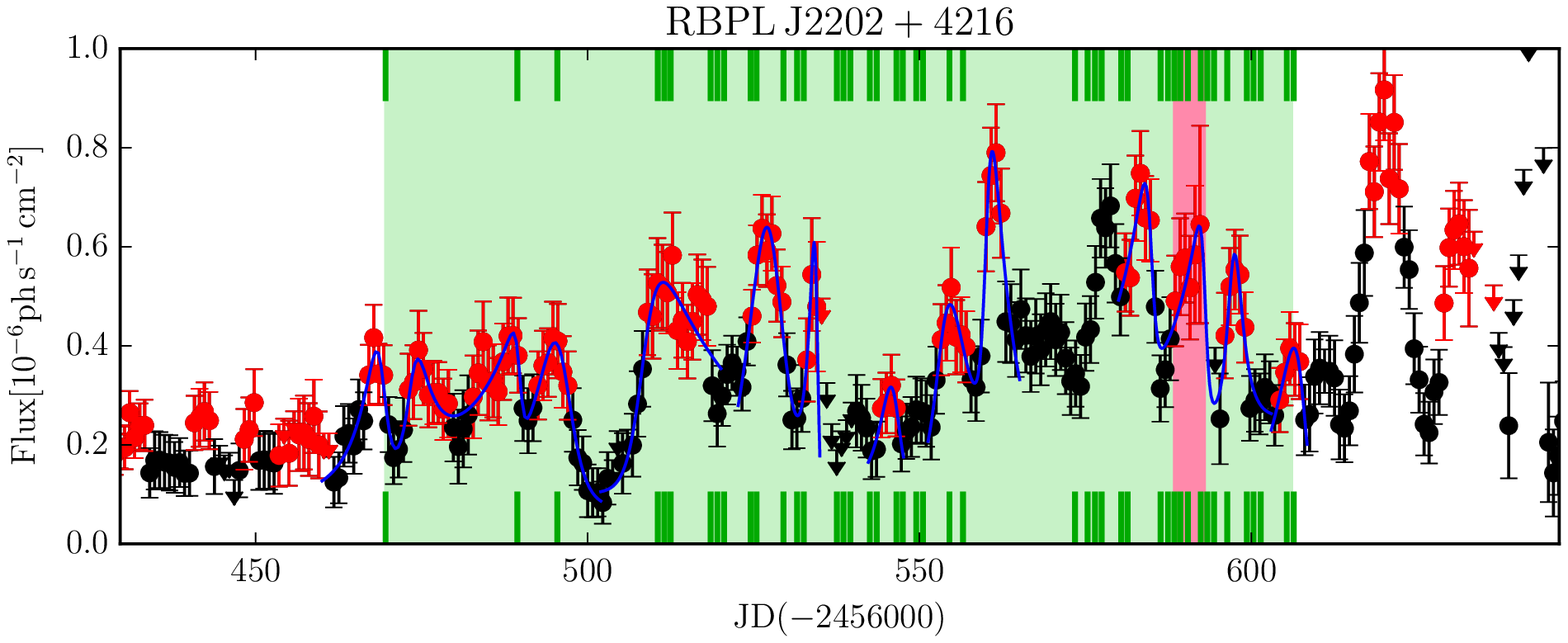}\\
 \includegraphics[width=0.494\textwidth]{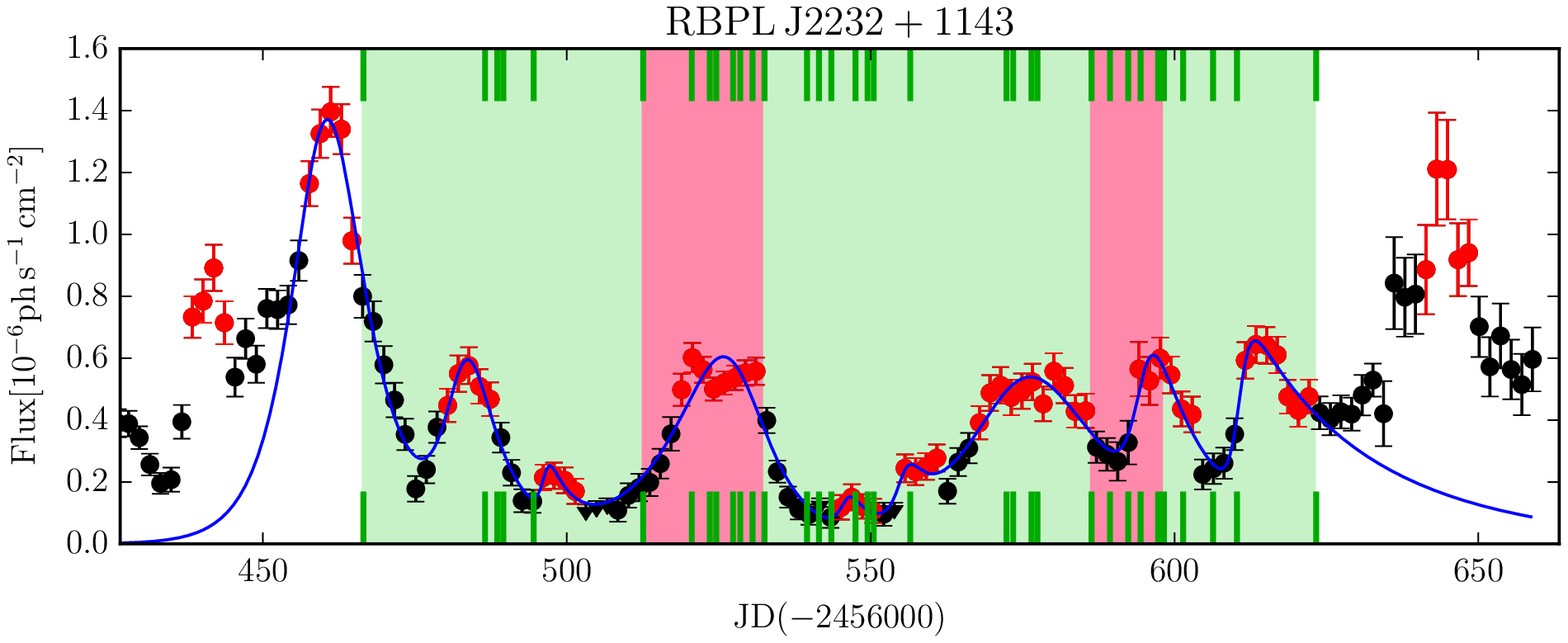}
 \includegraphics[width=0.494\textwidth]{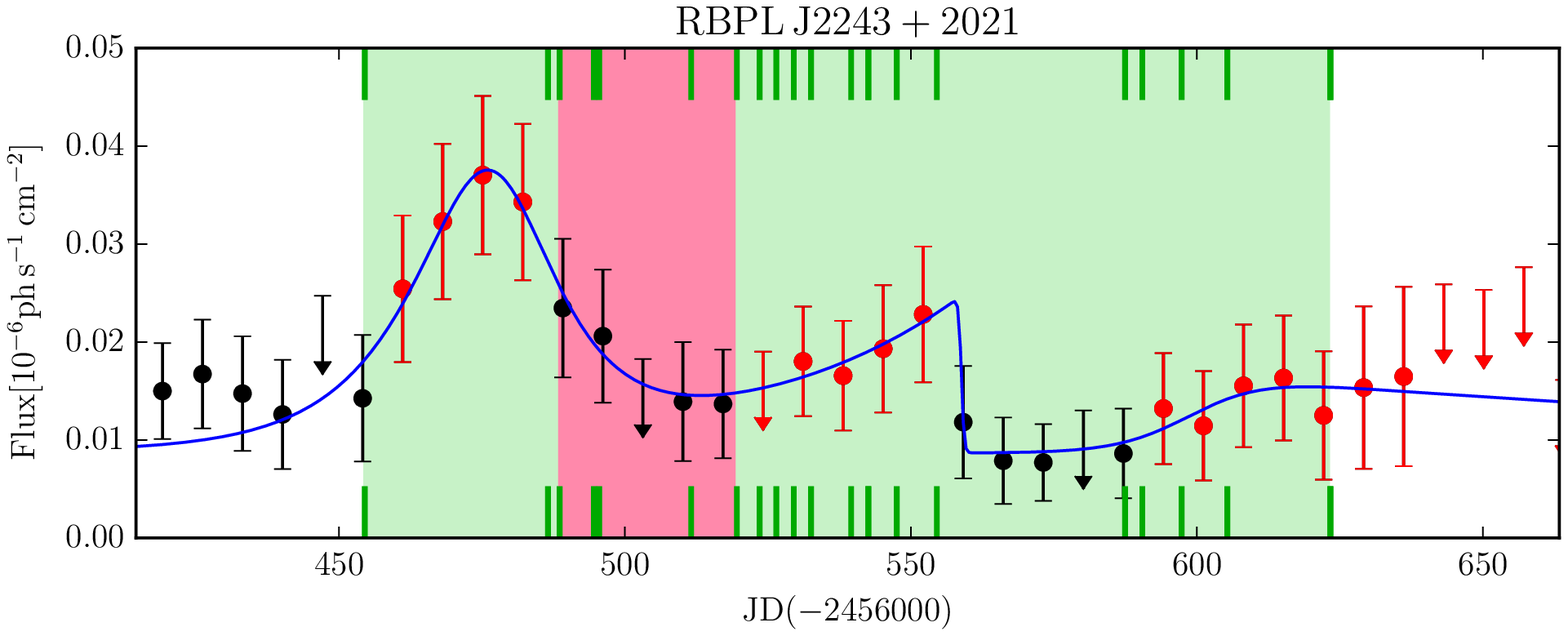}\\
\caption{Gamma-ray light curves of objects with detected rotations of EVPA during the 2013 {\em 
RoboPol} observing season. The season interval is marked by the green (light) area. The pink (dark) 
area shows duration of the rotation. Green ticks mark moments of our optical EVPA measurements. The 
red points (light grey in black and white renderings of the figures) indicate intervals identified 
as flares. The points are separated by $t_{\rm int}/4$. The blue line is the best least-square fit 
of Eq.~\ref{fit_func} to the data.}
\label{fig:rotations1}
\end{figure*}
\begin{figure*}
 \centering
 \includegraphics[width=0.481\textwidth]{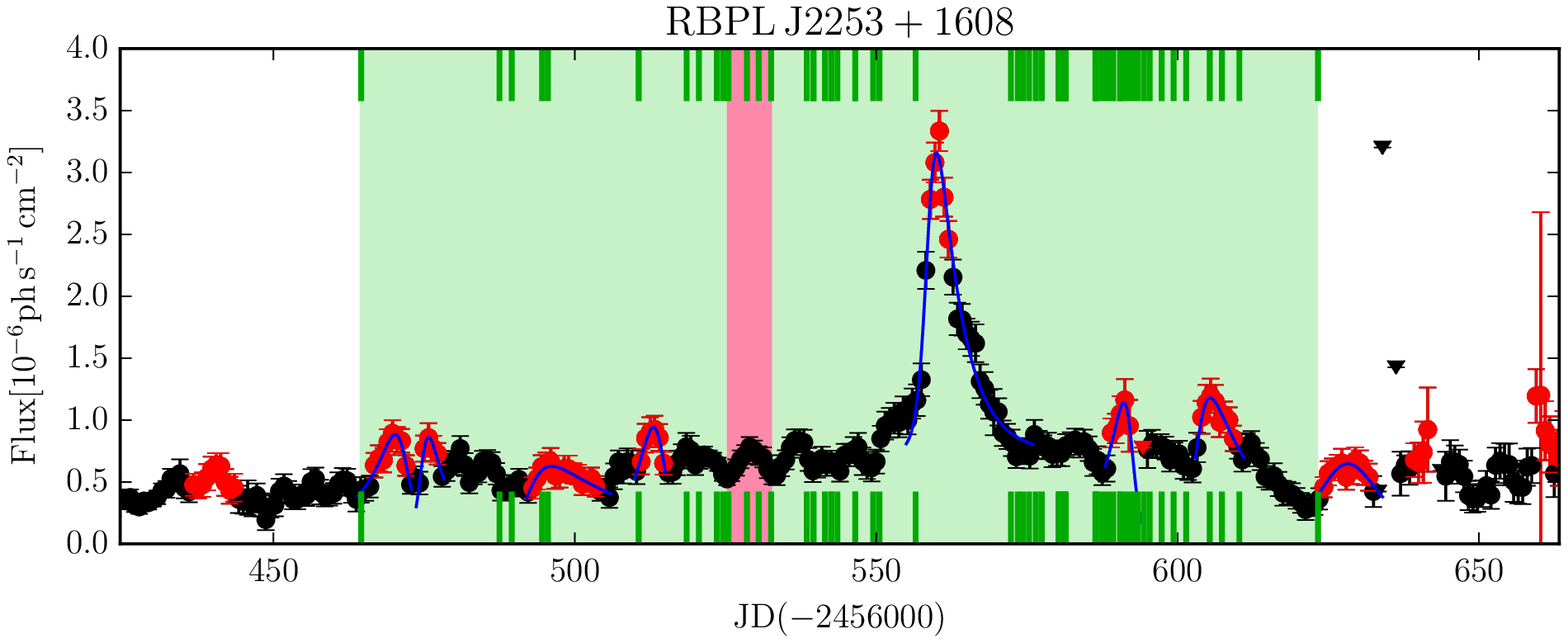}
 \includegraphics[width=0.481\textwidth]{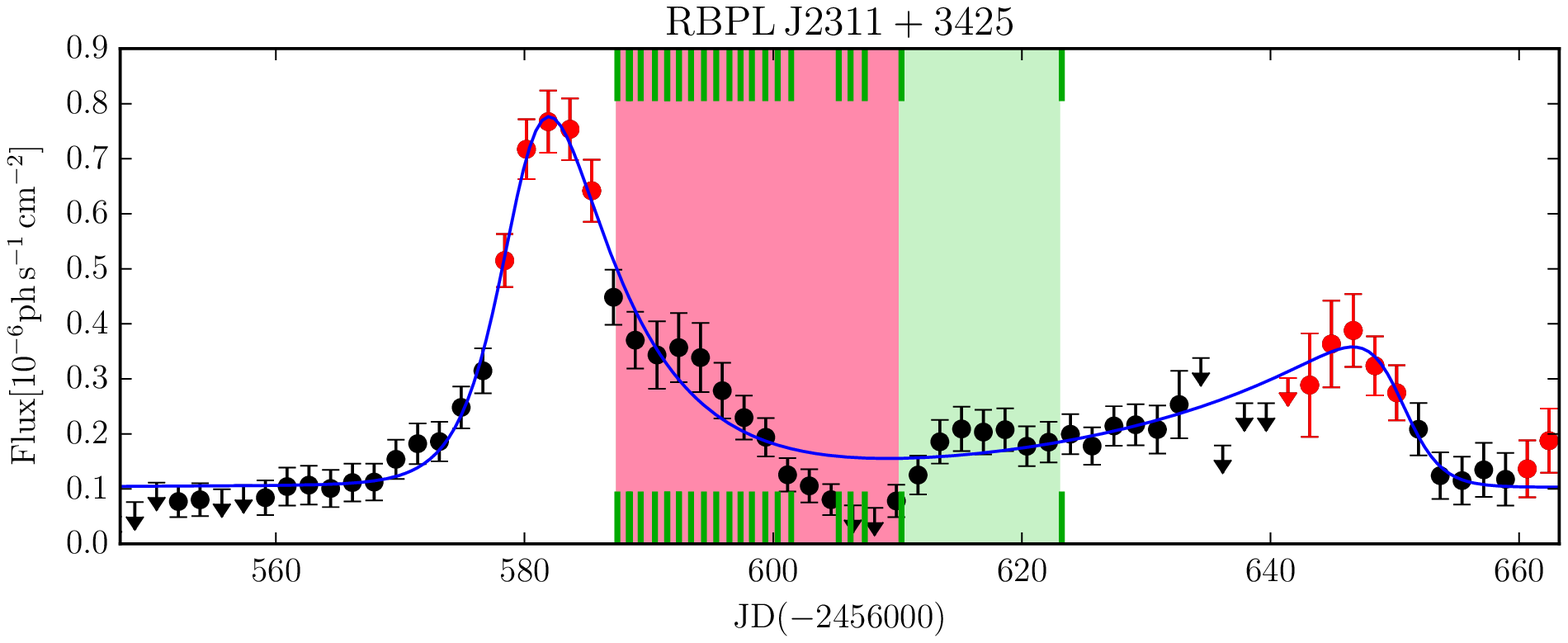}\\
 \contcaption{\label{fig:rotations1a}}
\end{figure*}

\begin{figure*}
\centering
  \includegraphics[width=0.481\textwidth]{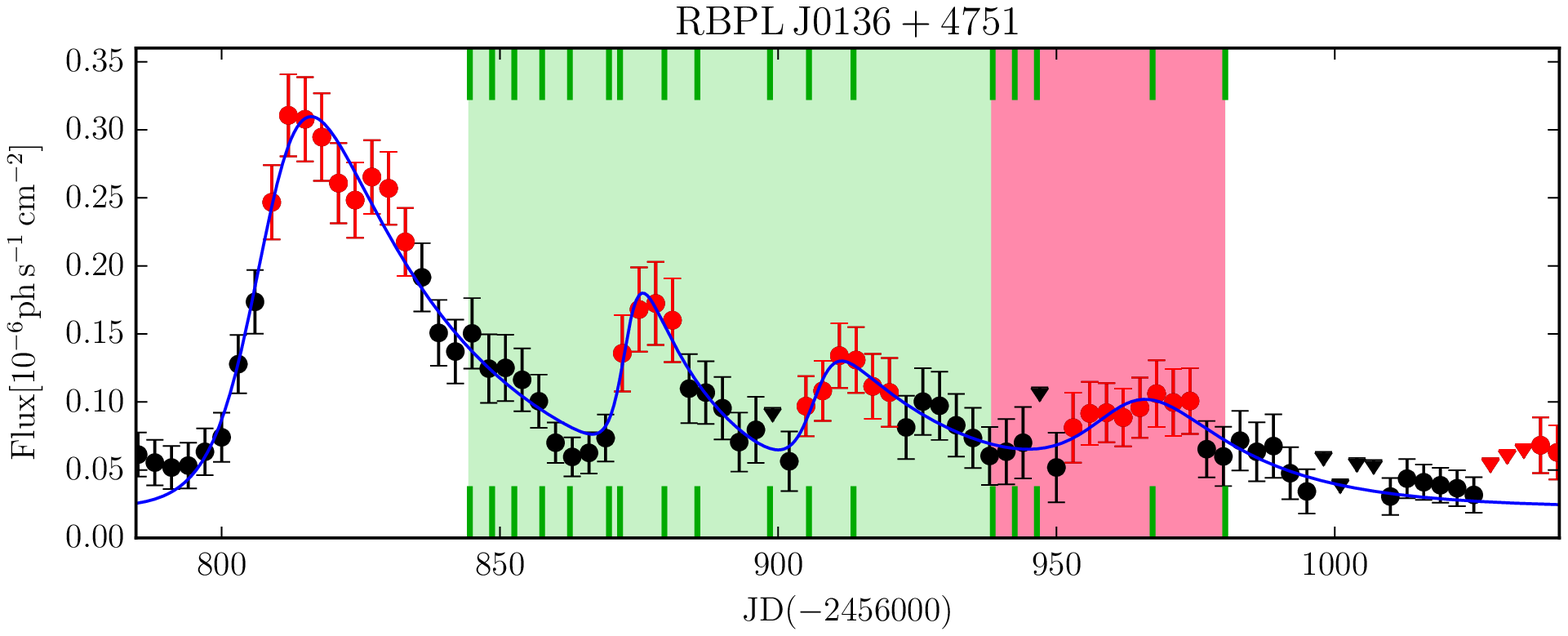}
  \includegraphics[width=0.481\textwidth]{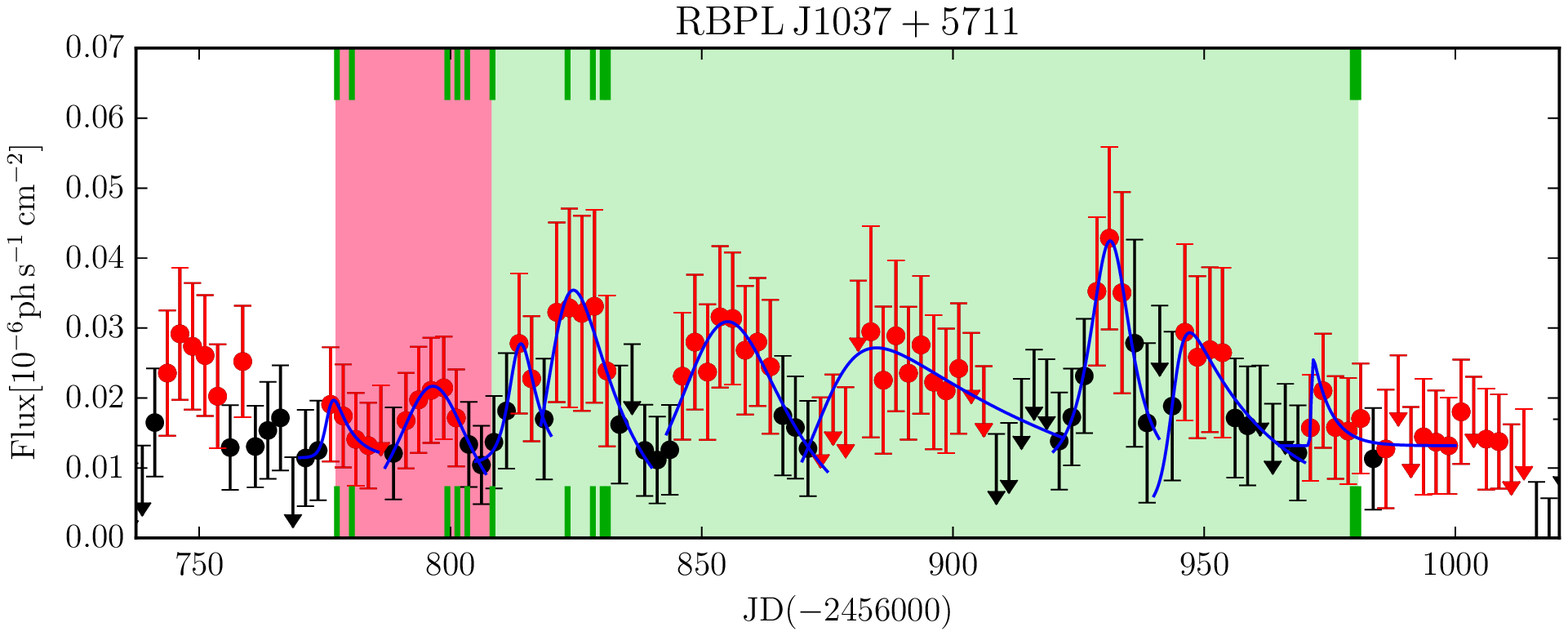}\\
  \includegraphics[width=0.481\textwidth]{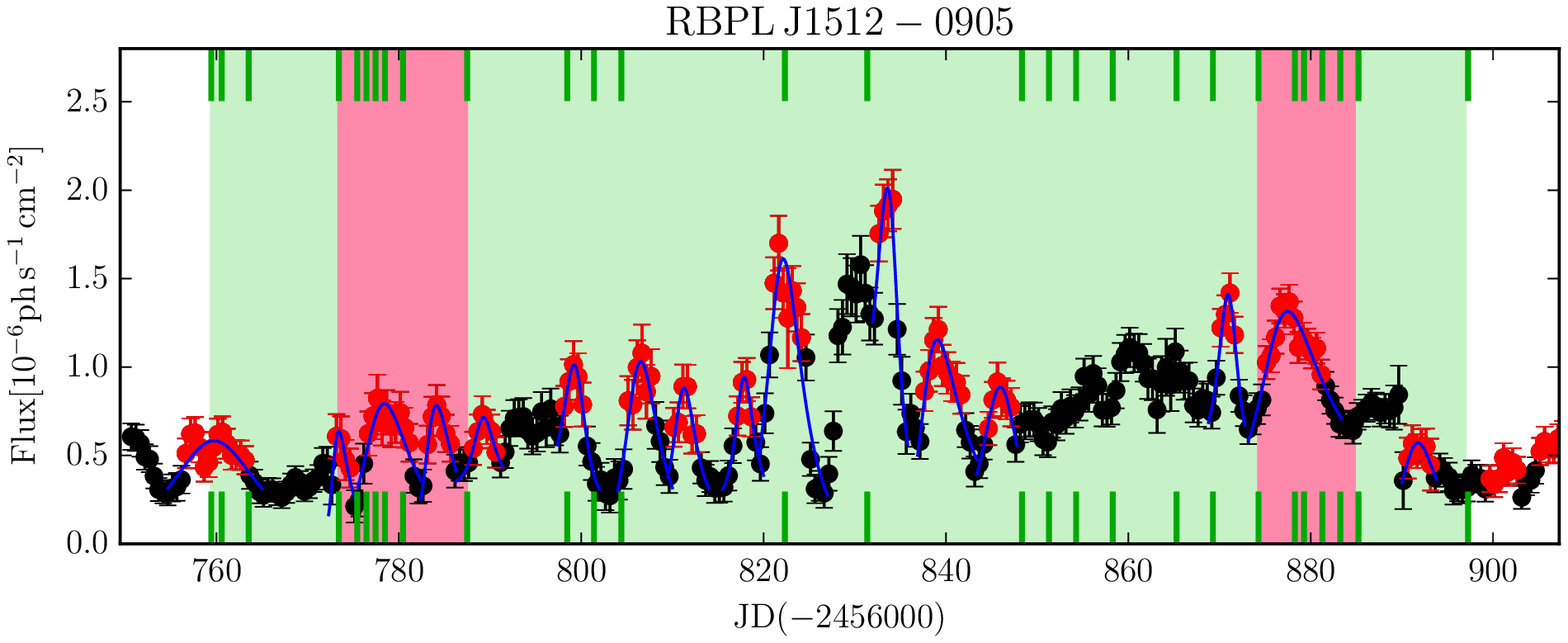}
  \includegraphics[width=0.481\textwidth]{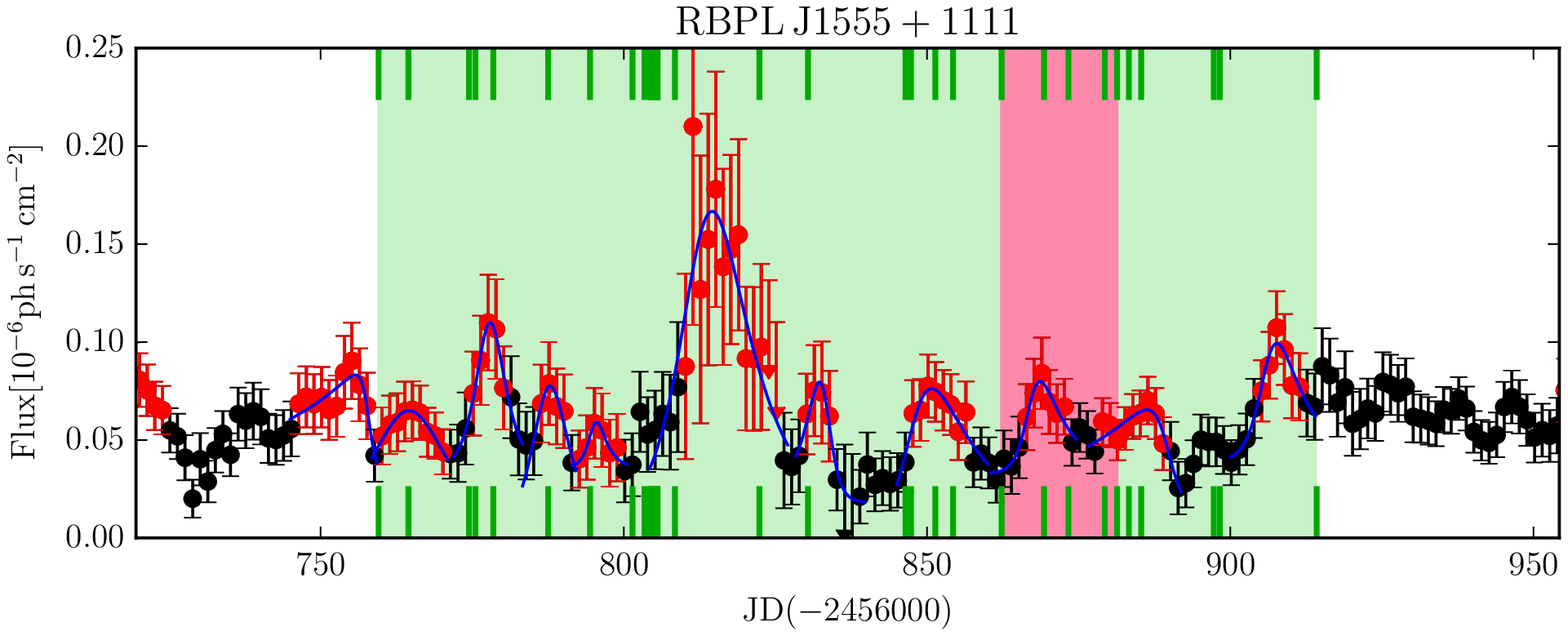}\\
  \includegraphics[width=0.481\textwidth]{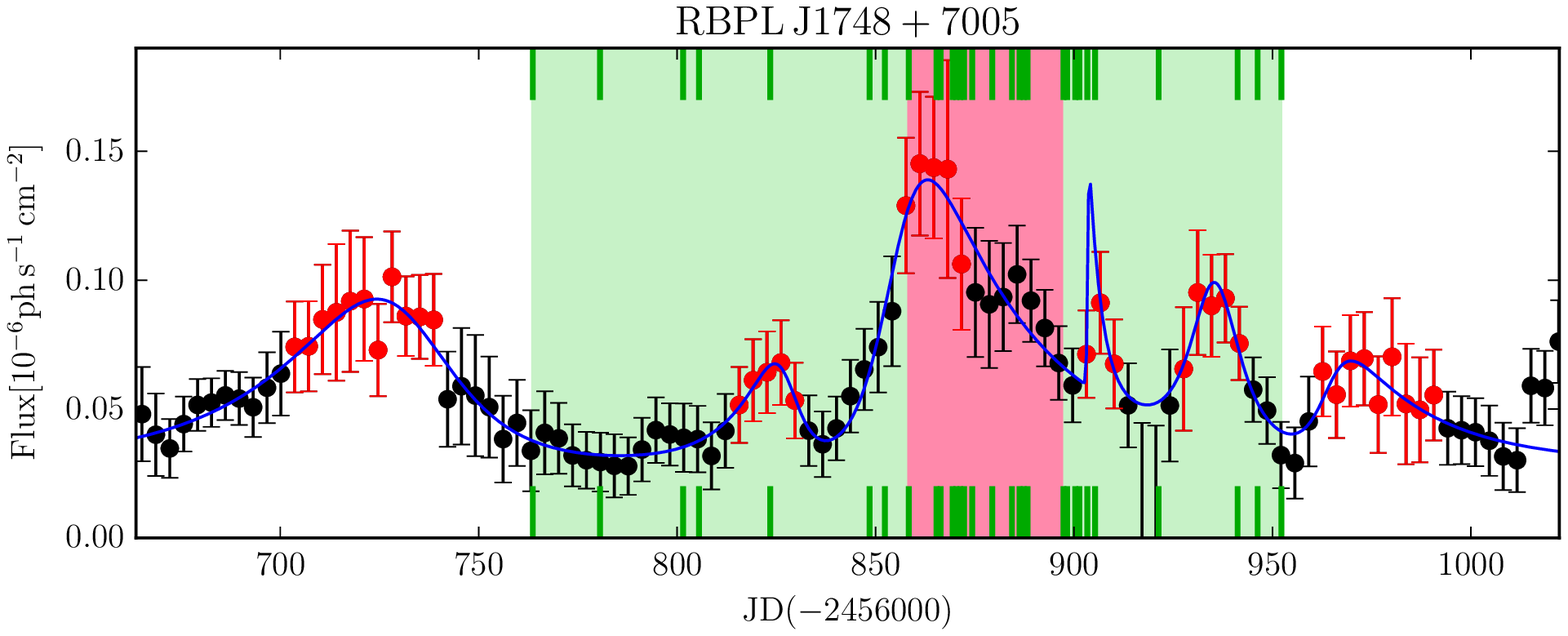}
  \includegraphics[width=0.481\textwidth]{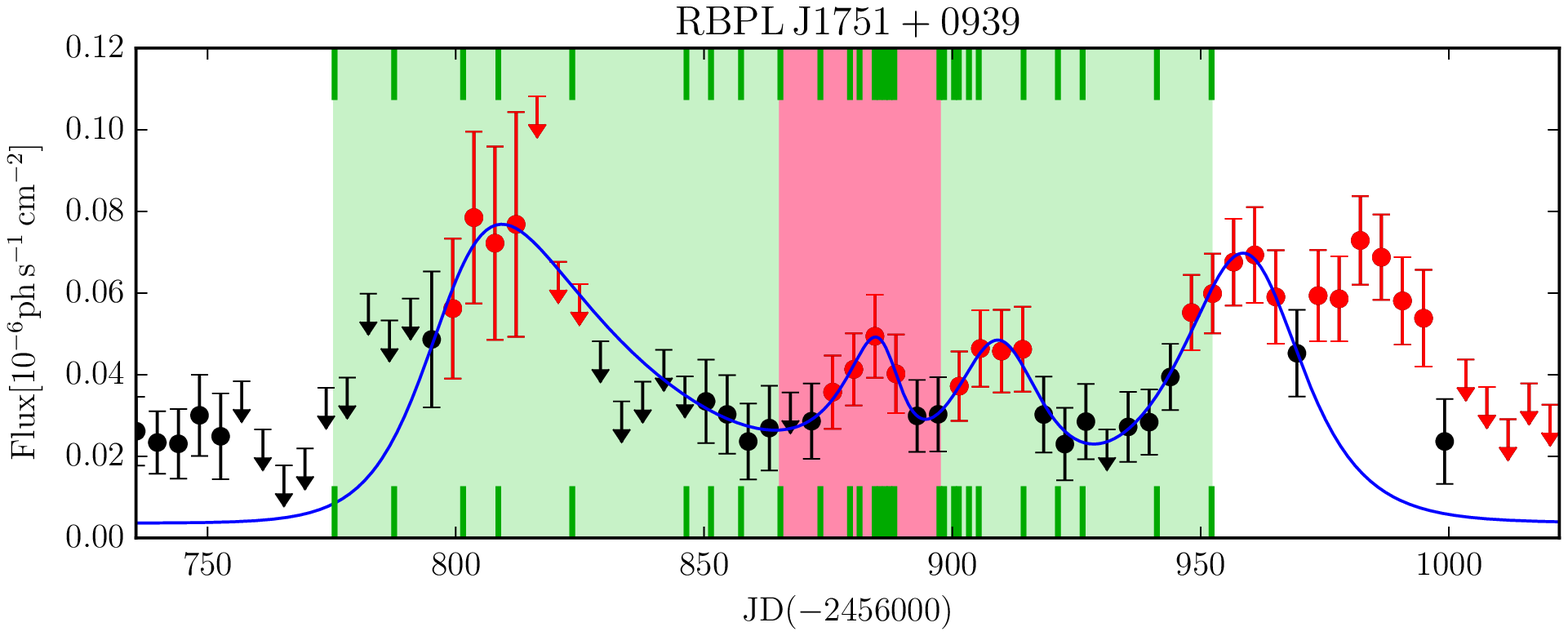}\\
  \includegraphics[width=0.481\textwidth]{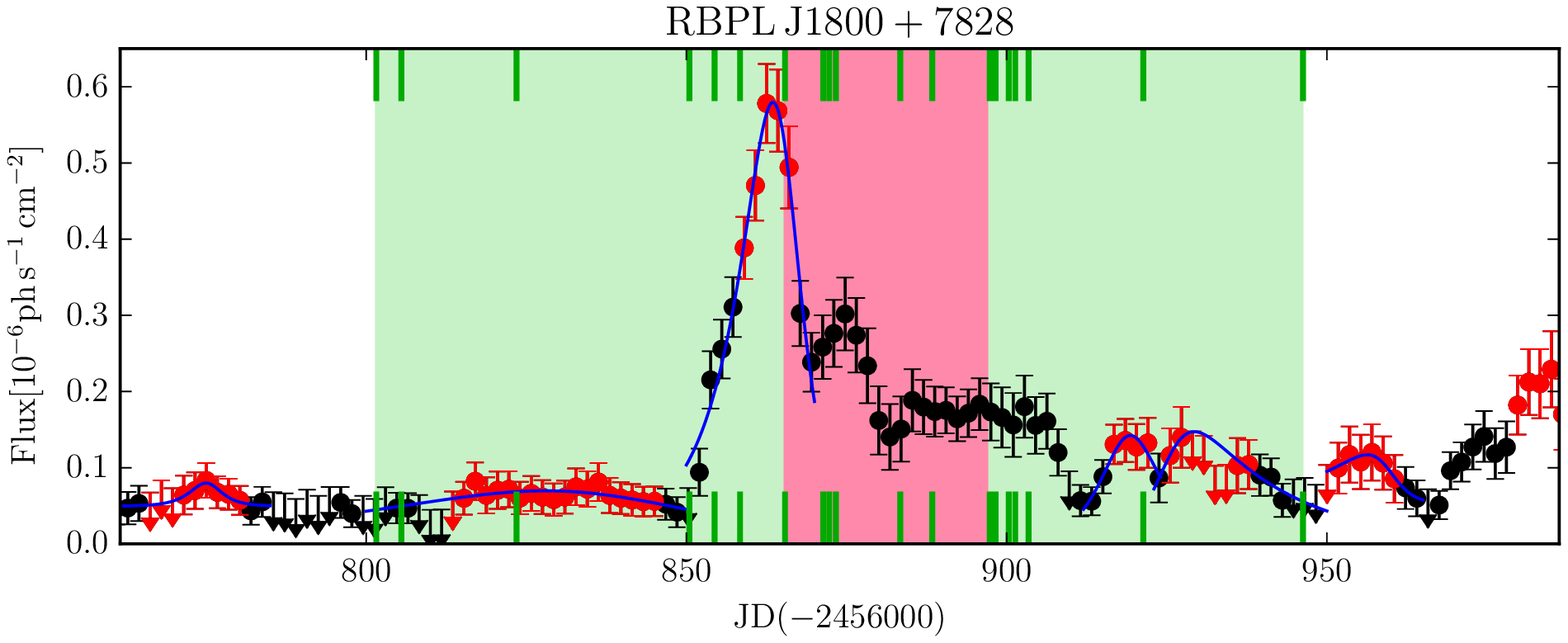}
  \includegraphics[width=0.481\textwidth]{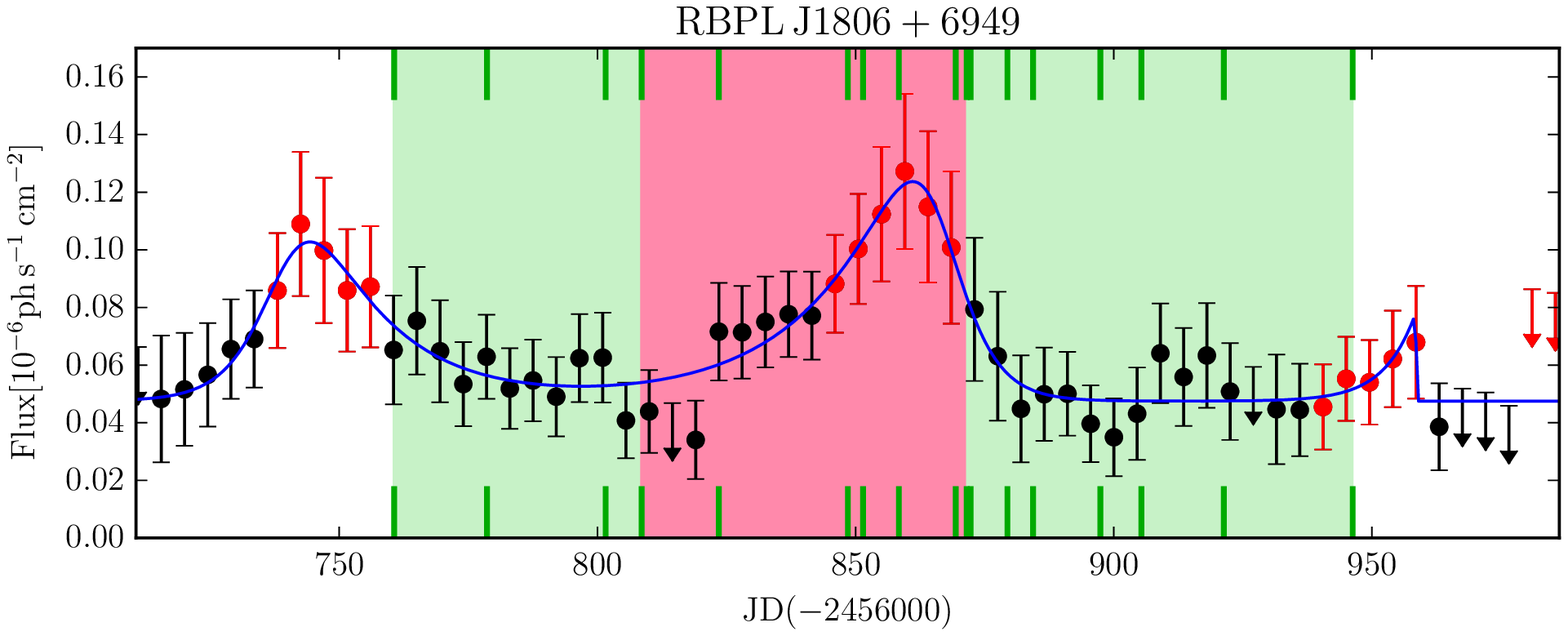}\\
  \includegraphics[width=0.481\textwidth]{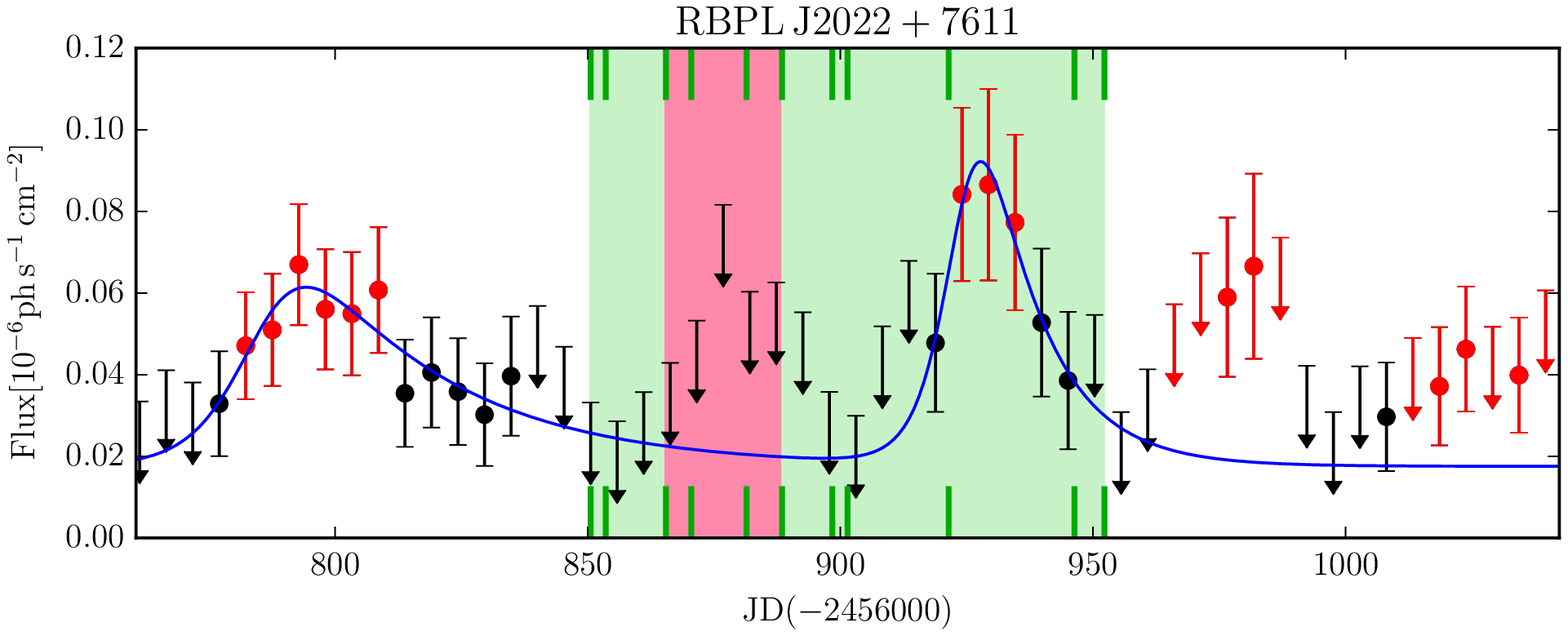}
  \includegraphics[width=0.481\textwidth]{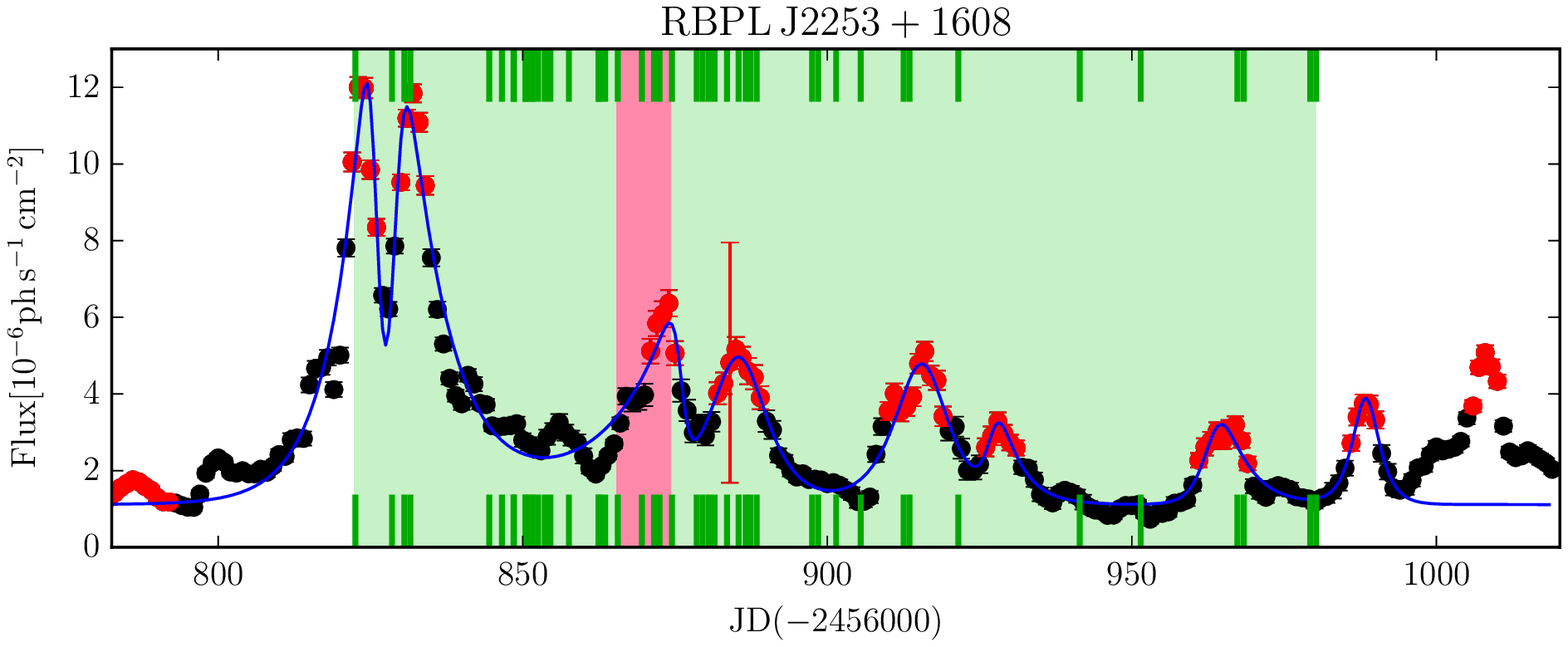}\\
  \caption{Same as Fig. \ref{fig:rotations1} for the 2014 observing season.}
  \label{fig:rotations2}
\end{figure*}

\begin{figure*}
  \centering
  \includegraphics[width=0.494\textwidth]{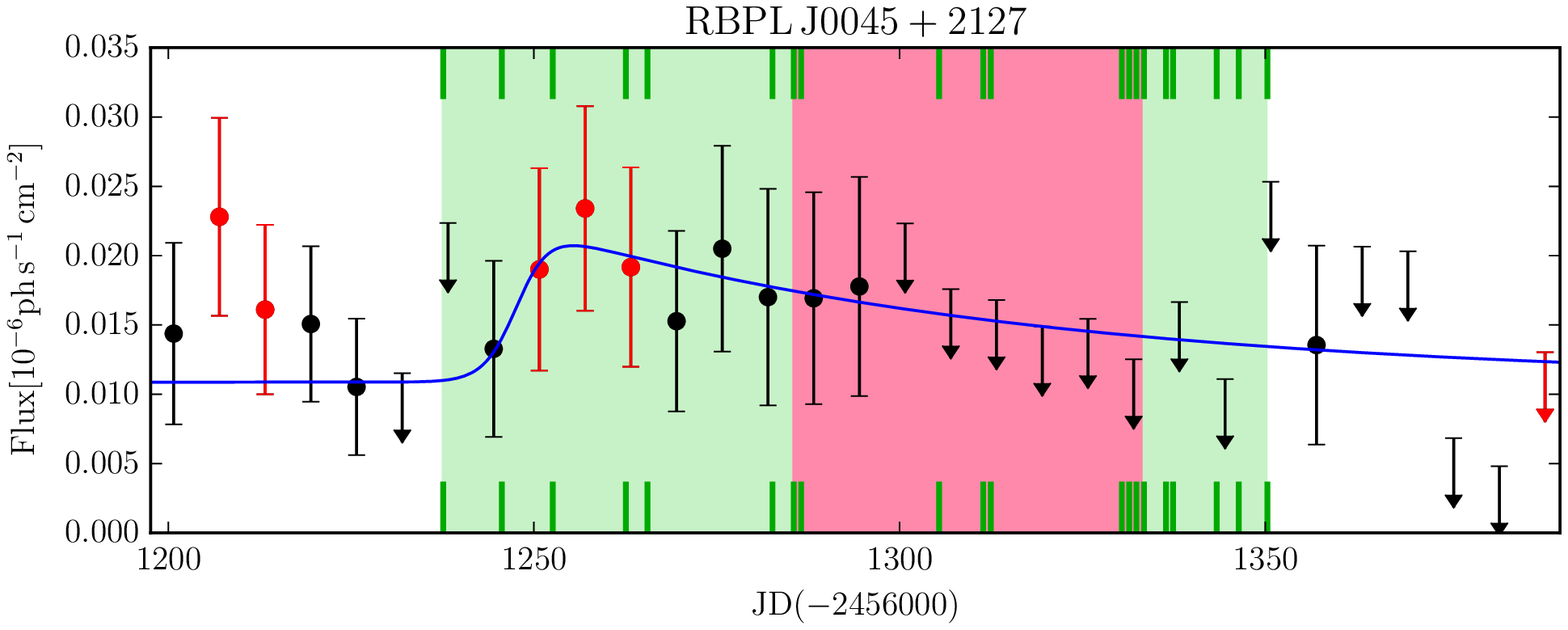}
  \includegraphics[width=0.494\textwidth]{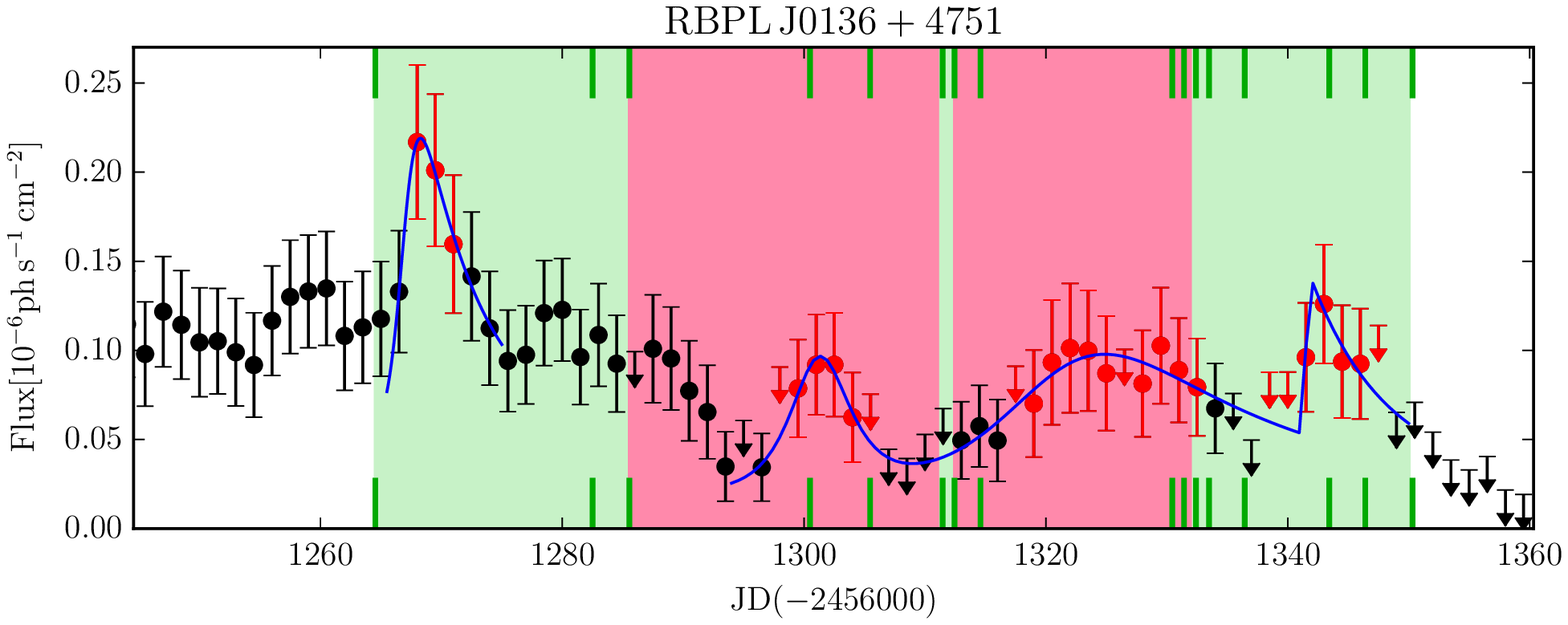}\\
  \includegraphics[width=0.494\textwidth]{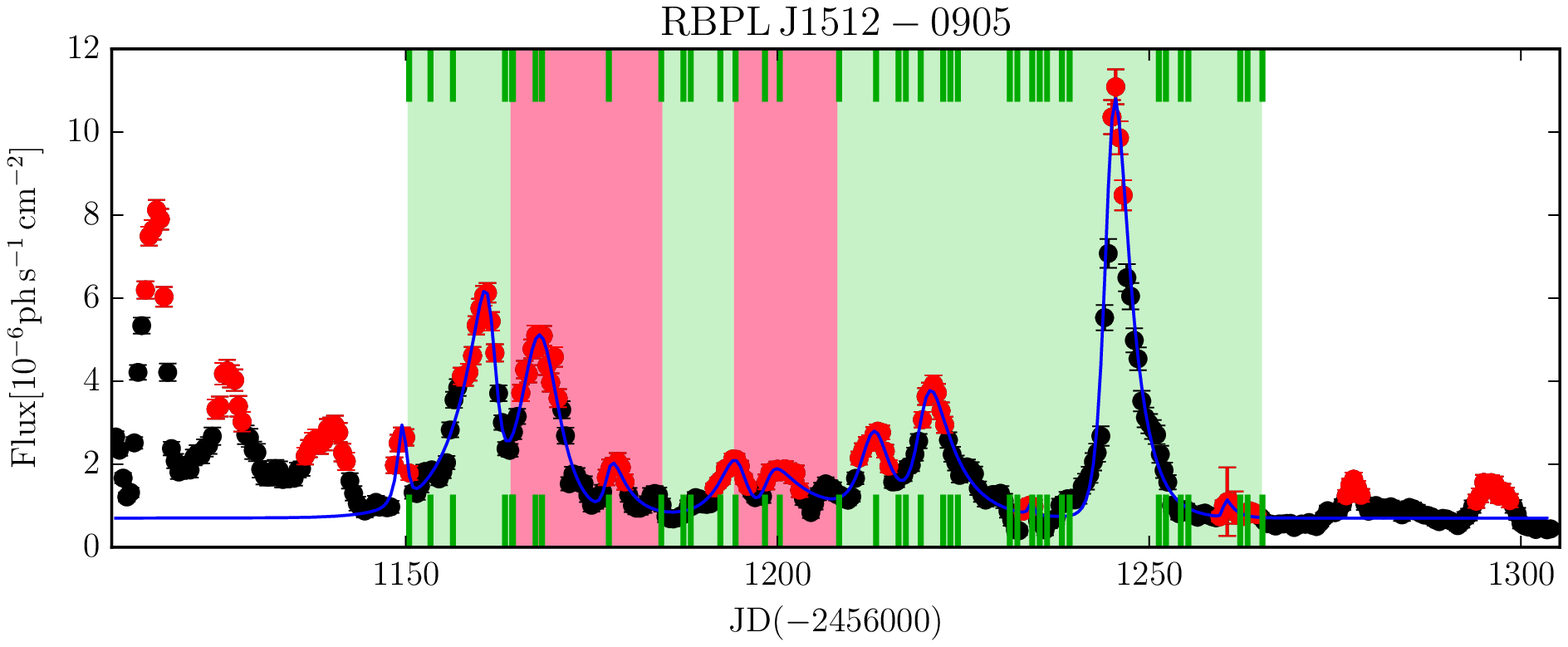}
  \includegraphics[width=0.494\textwidth]{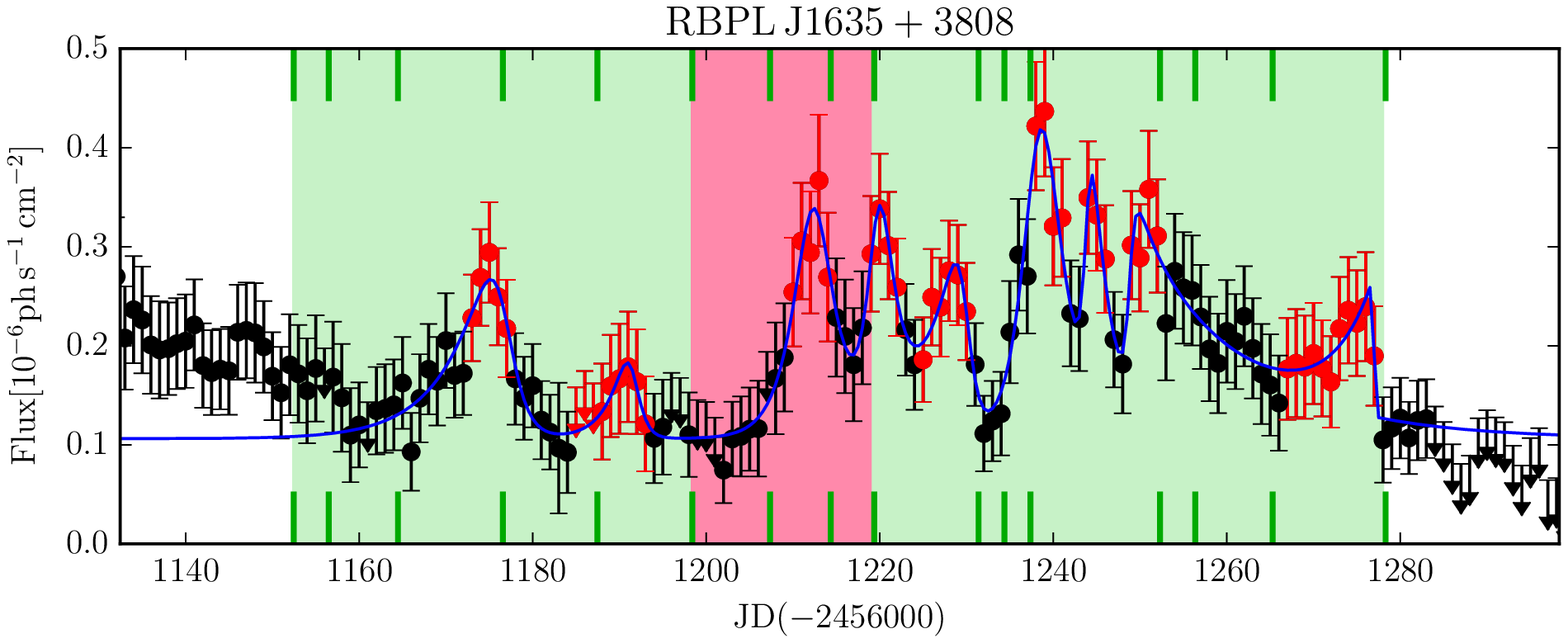}\\
  \includegraphics[width=0.494\textwidth]{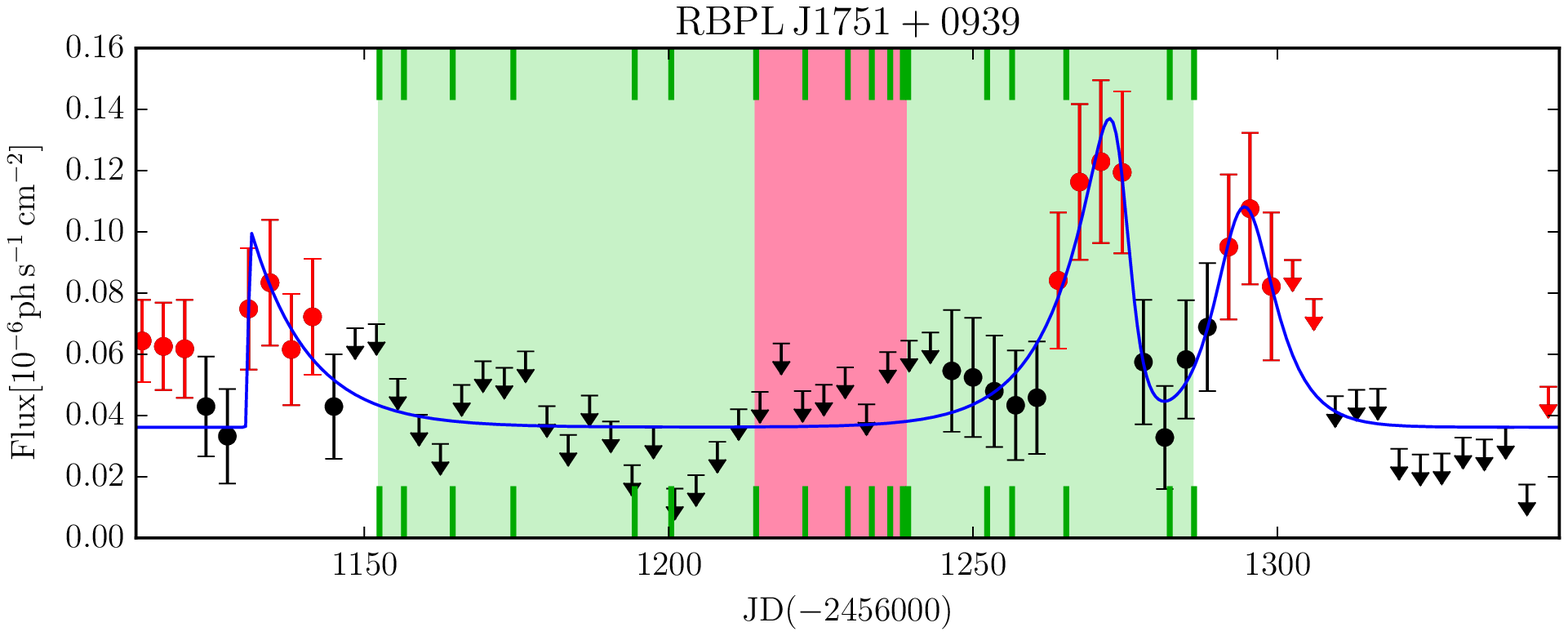}
  \includegraphics[width=0.494\textwidth]{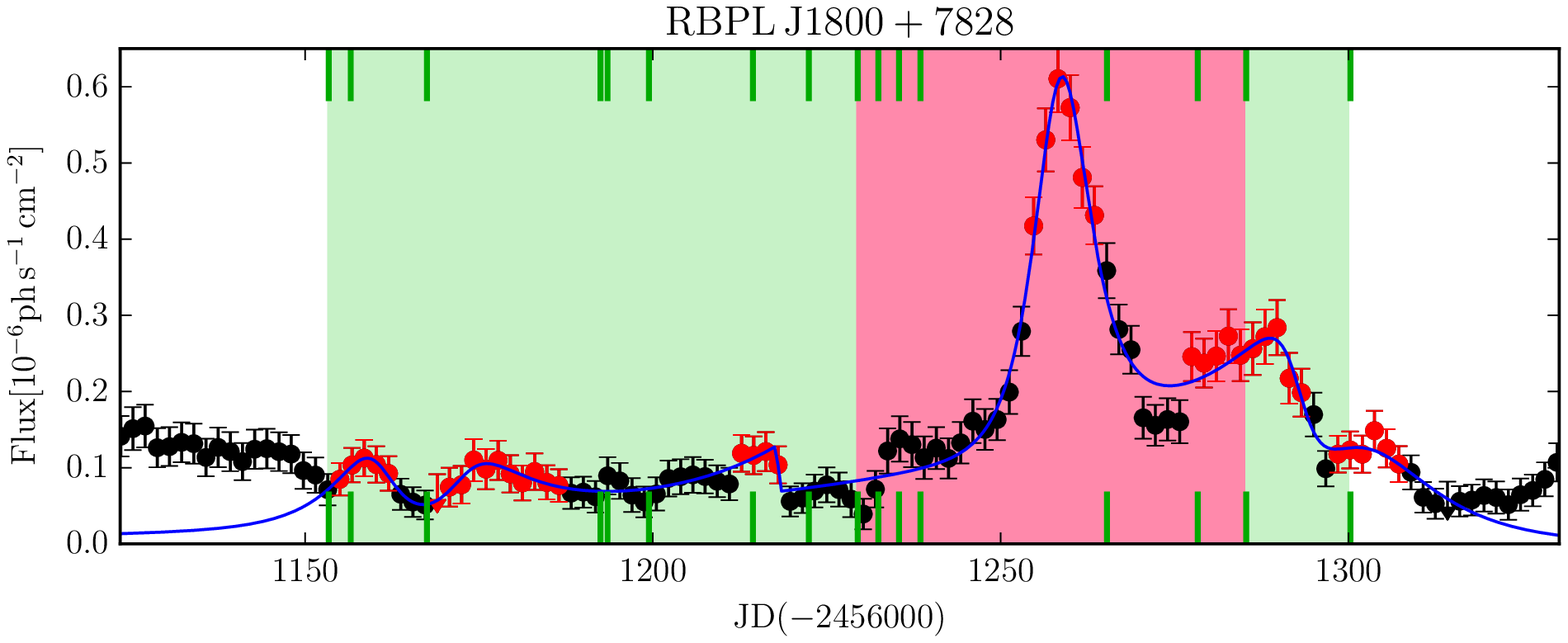}\\
  \includegraphics[width=0.494\textwidth]{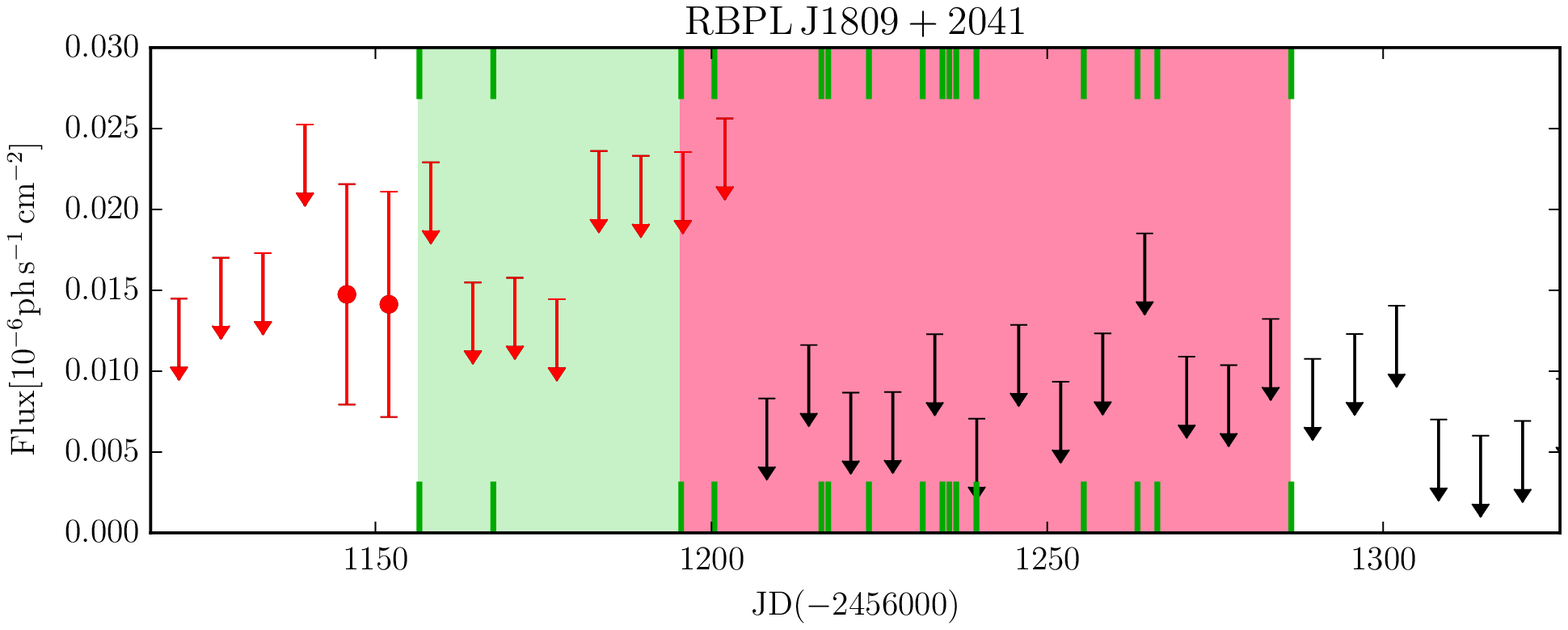}
  \includegraphics[width=0.494\textwidth]{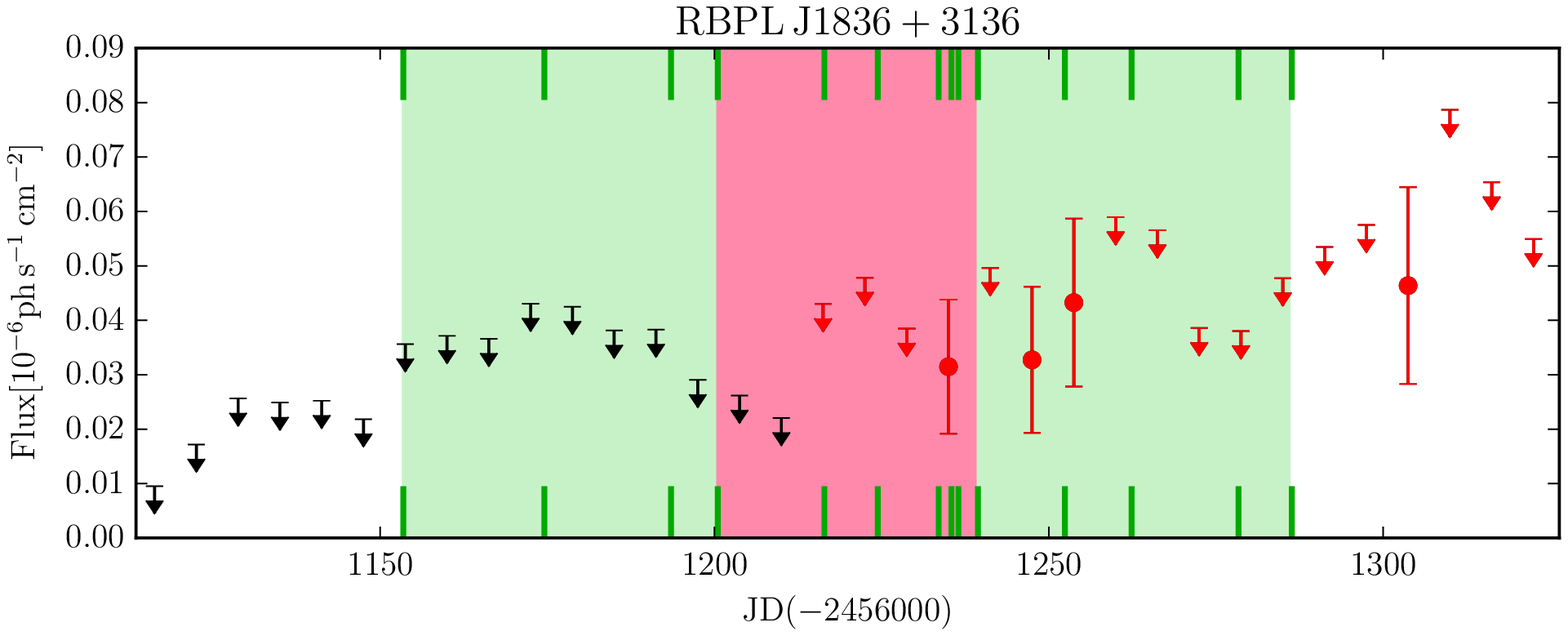}\\
  \includegraphics[width=0.494\textwidth]{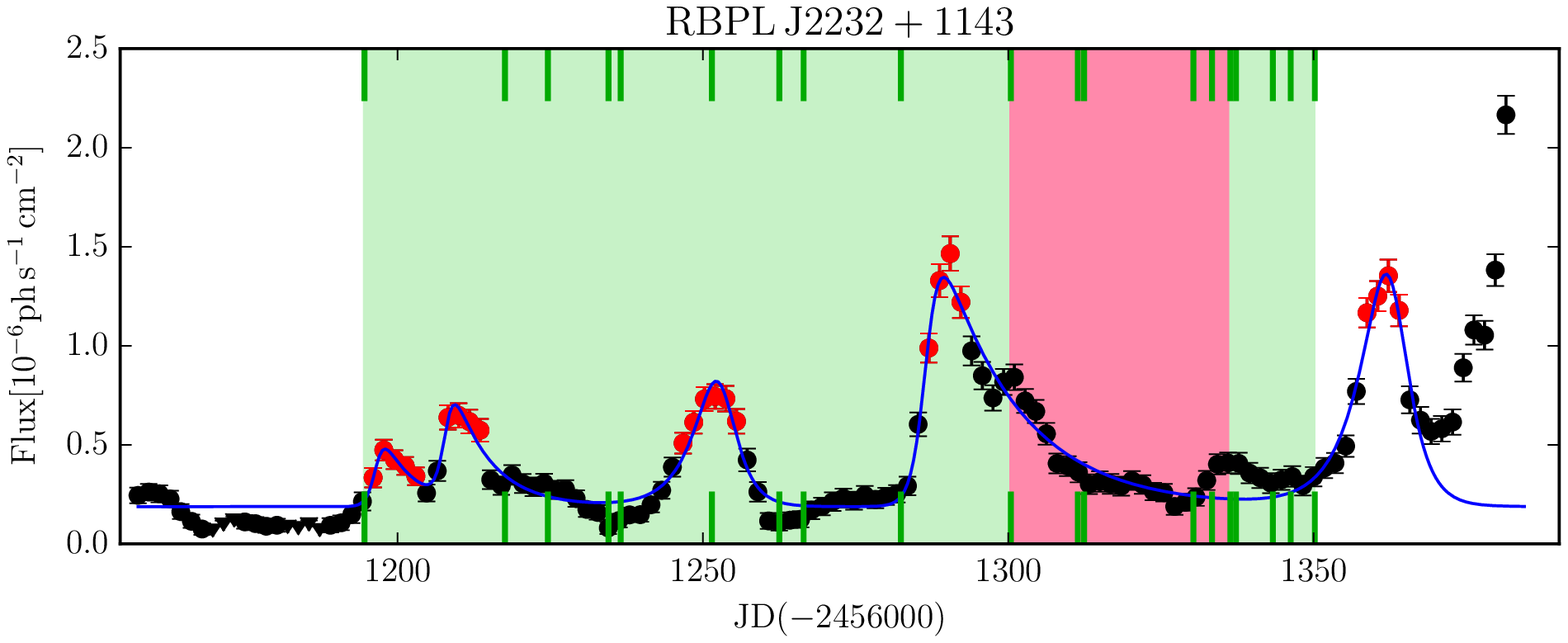}
  \includegraphics[width=0.494\textwidth]{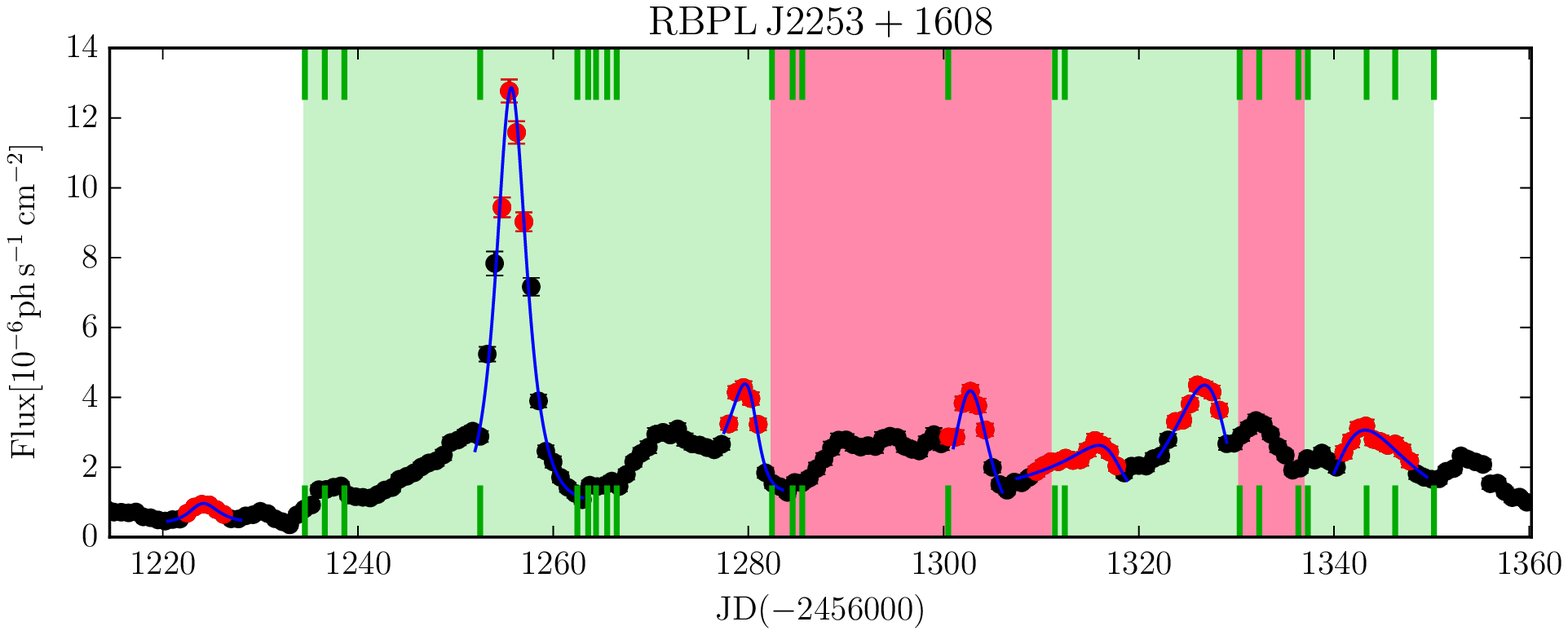}\\
  \caption{Same as Fig. \ref{fig:rotations1} for the 2015 observing season.}
  \label{fig:rotations3}
\end{figure*}

However, we must examine the statistical significance of this apparent difference. The mean time 
lag, $\overline{\tau_{\rm corr}}$, for the high and low amplitude samples is 4.2 and $-1.9$ d. 
According to the Student's t-test the null-hypothesis of identical mean values is accepted 
($p\text{-value} = 0.21$). A similar conclusion holds for absolute values of $\tau_{\rm corr}$. The 
standard deviation of time lags, $\sigma_{\tau_{\rm corr}}$, for the high and low amplitude samples 
is 6.1 d and 11.8 d. According to tests for equality of variances by \cite{Levene1960} and 
\cite{Bartlett1937}, we cannot reject the null-hypothesis that the standard deviations are equal 
($p\text{-value} = 0.51$ and $0.11$). Based on these results, we conclude that $\overline{\tau_{\rm 
corr}}$ and $\sigma_{\tau_{\rm corr}}$ for the high and low amplitude flares are consistent with 
being equal. Hence, the apparent distinction between high and low amplitude flares in 
Fig.~\ref{fig:TLvsAmplobs} is statistically insignificant.

\subsection{Are all time lags consistent with zero?} 
\label{subsec:ampl_tl_explan}

The results from the analysis in section \ref{subsec:obs_ampl_tl} suggest that $\overline{\tau_{\rm 
corr}}$ is the same for the high and low amplitude gamma-ray flares. Although the distributions of 
$\tau_{\rm corr}$ in the two groups are sparsely sampled, it appears possible that the two are 
identical. The larger scatter of $\tau_{\rm corr}$ in low amplitude flares may be attributed to 
experimental noise, the effects of which are more pronounced in the case of light curves with low 
photon counts.

To investigate this effect for each of the gamma-ray light curves we computed the ratio $N_{\rm 
up}/N_{\rm det}$, where $N_{\rm det}$ is the number of points  with detected photon flux (i.e., $TS 
> 10$) and $N_{\rm up}$ is the number of determined upper limits. Both numbers only refer to the 
time range corresponding to the {\em RoboPol} observing season of each blazar.
\begin{figure}
 \centering
 \includegraphics[width=0.42\textwidth]{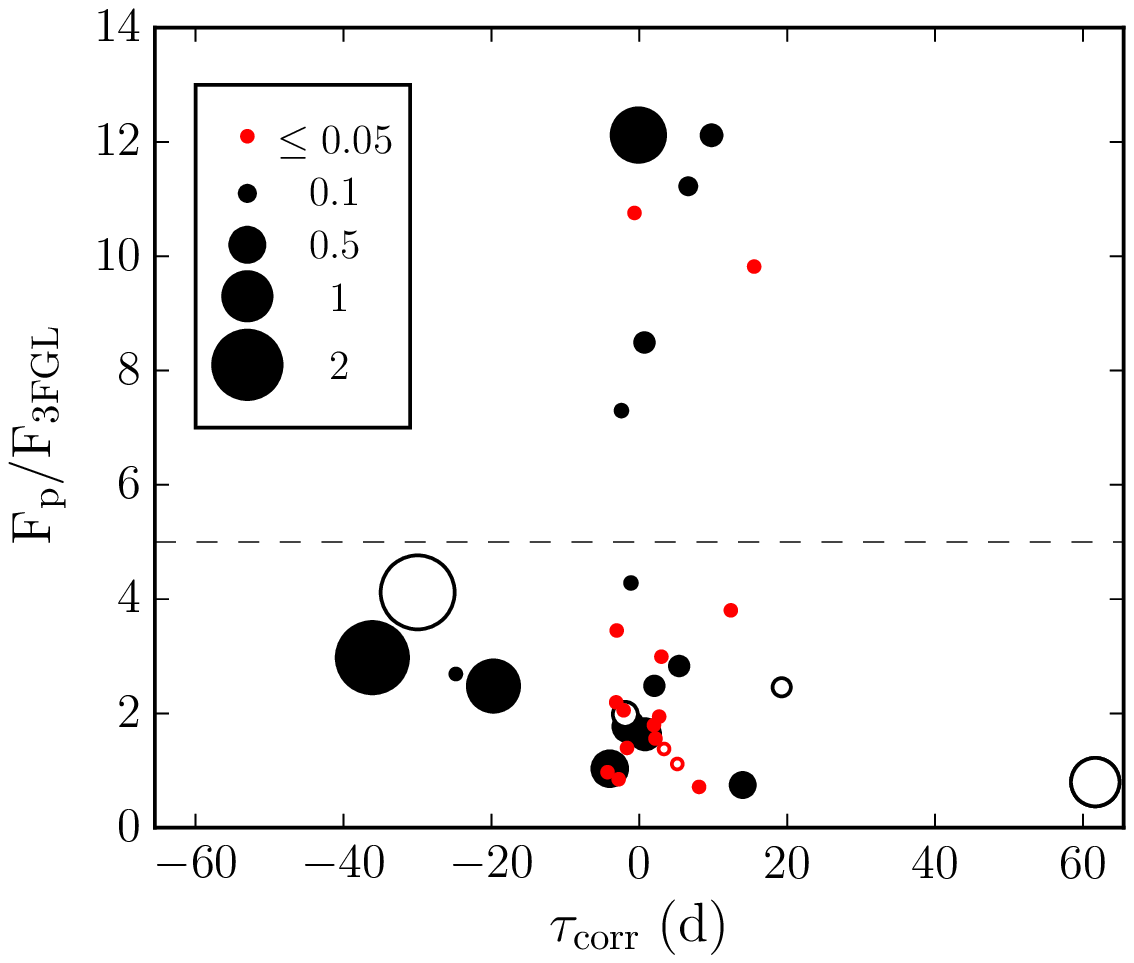}
 \caption{Time lags, $\tau_{\rm corr}$, vs. normalized gamma-ray flare amplitude, $F_{\rm p}/F_{\rm 
3FGL}$. The size of the points indicates $N_{\rm up}/N_{\rm det}$ -- the fraction of the {\em 
RoboPol} observing season when the photon flux of the rotator was lower than the \textit{Fermi} LAT 
detection limit. Empty symbols represent events in blazars with uncertain or unknown redshift.}
 \label{fig:TLvsAmplDuty}
\end{figure}
In Fig.~\ref{fig:TLvsAmplDuty} we show the distribution on the $\tau_{\rm corr}$ -- $F_{\rm 
p}/F_{\rm 3FGL}$ plane, where the size of points is proportional to $N_{\rm up}/N_{\rm det}$. 
This figure suggests that points with larger time lags between rotations and flares tend 
to also have larger fraction of non-detections in the gamma-ray photon flux curves. In fact, the 
Pearson's correlation coefficient for $|\tau_{\rm corr}|$ and $N_{\rm up}/N_{\rm det}$ is 0.58 
($p\text{-value} = 2\times10^{-4}$). As a result, the time lags in some cases may not be real, as 
the gamma-ray flare truly associated with the EVPA is not sampled due to low statistics of 
data points with significant flux detection.

Furthermore, we compared current data with those used in Paper~I, where the \textit{Fermi}-LAT 
Pass 7 data were analysed with the previous version of Science Tools v9r33p0, and the photon fluxes 
were normalized with average fluxes of blazars from the 2FGL catalogue \citep{Nolan2012}.
\begin{figure}
 \centering
 \includegraphics[width=0.46\textwidth]{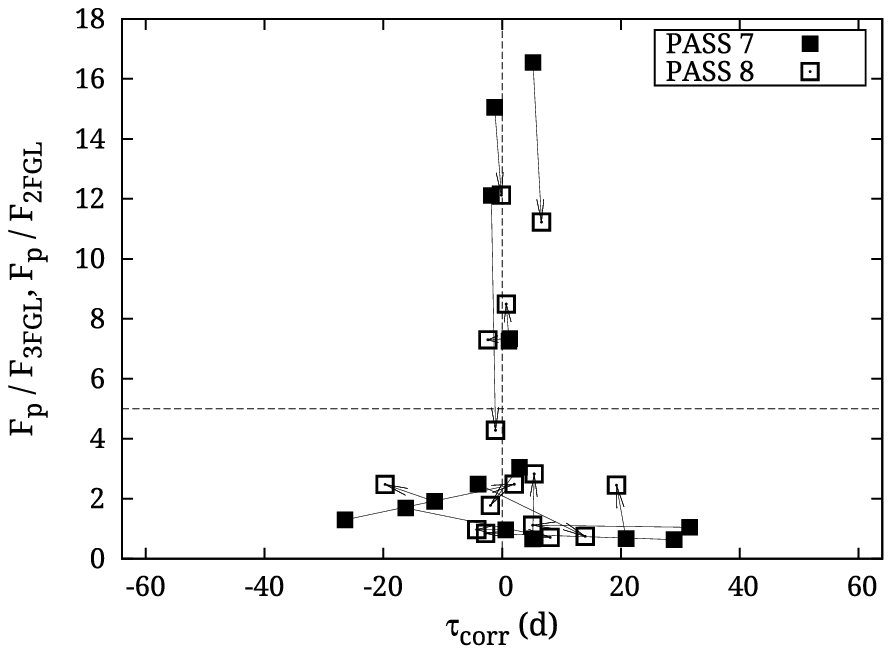}
 \caption{Time lags, $\tau_{\rm corr}$, vs. normalized gamma-ray flare amplitude, $F_{\rm p}/F_{\rm 
3FGL}$ (open squares) and $F_{\rm p}/F_{\rm 2FGL}$ (filled squares), for the 2013 season events. 
For direct comparison with Paper~1 two other seasons are omitted.}
 \label{fig:TLvsAmplP7P8}
\end{figure}
The difference between 2FGL and 3FGL is that the later has a number of improvements in the
analysis. For instance, it uses better model of the diffuse Galactic and isotropic emissions as well
as more accurately characterized instrument response functions. Also the 3FGL catalogue incorporates
data from twice longer time span than 2FGL. However, we did not find any systematic difference in
photon fluxes for our sample of rotators between the two catalogues. Moreover, these values are 
strongly correlated ($r=0.998$) and the slope of the best fit line in the $F_{\rm 3FGL}$ vs 
$F_{\rm 2FGL}$ plane is consistent with unity. Fig.~\ref{fig:TLvsAmplP7P8} shows the difference
between Pass 7 and Pass 8 data normalized with 2FGL and 3FGL average photon fluxes. The filled 
squares represent data from Fig.~8 in Paper~I and the empty squares are the data for corresponding 
events from the current work. The figure shows that high amplitude flares preferentially changed the 
relative amplitude between the two versions of the analysis, while low amplitude events mostly 
changed the time lag. The mean difference of the time lag for low amplitude events in the two data 
sets in Fig.~\ref{fig:TLvsAmplP7P8} is 16.3 d. It is close to $\sigma_{\tau_{\rm corr}}=18$ d of the 
low amplitude events in Fig.~\ref{fig:TLvsAmplobs}.
This implies that unaccounted uncertainties mostly affect amplitudes and time lags for high and low  
amplitude events, respectively. In other words, if a gamma-ray flare is strong enough, the position
of its peak (and hence $\tau_{\rm corr}$) is not affected by any experimental uncertainties that 
have not been taken into account by the data analysis procedure. On the other hand, if an EVPA 
rotation is intrinsically linked with a gamma-ray flare of a small amplitude, then the 
determination of their time lag will be difficult due to the low-counts statistics, which may 
prevent the accurate determination of the flare peak.

Based on these arguments, we conclude that the higher apparent spread of time lags 
between rotations and low amplitude gamma-ray flares compared to that of high amplitude events is
not necessarily a real effect. Rather, it could be caused by insufficient sampling of the gamma-ray 
photon flux curves (due to low photon statistics) and the subsequent uncertainties in the 
determination of the flare parameters. Therefore, Fig.~\ref{fig:TLvsAmplobs} cannot be considered 
as evidence of two types of EVPA rotations. Our analysis indicates that all EVPA rotations are 
related to gamma-ray flares with $\overline{\tau_{\rm corr}}=1.5 \pm 3.1$ d, which (given the 
uncertainties in our estimate of the beginning and end of a rotation) is fully consistent with zero 
values i.e., {\em all EVPA rotations could be simultaneous with gamma-ray flares}.

\subsection{Are the time lags random?} \label{subsec:CDFs_TL}

\begin{figure}
 \centering
 \includegraphics[width=0.42\textwidth]{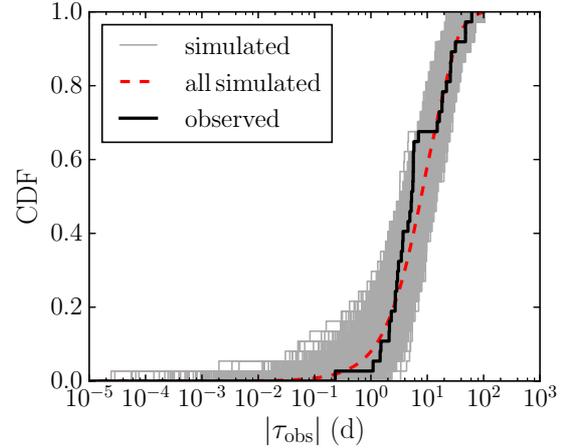}
\caption{CDFs of the time lags between the EVPA rotations and $t_{\rm p}$ of the closest gamma-ray 
flares for the main sample rotators. Black line -- observed time lags, thin grey lines -- $10^4$ 
simulated values for the whole sample of rotations (see text for details).}
 \label{fig:CDF_delays}
\end{figure}

The result of section \ref{subsec:ampl_tl_explan} suggests that EVPA rotations happen simultaneously 
with gamma-ray flares. However, most of the gamma-ray light curves in Fig.~\ref{fig:rotations1} - 
\ref{fig:rotations3} show many flares. Therefore, it is possible that the simultaneity between 
flares and rotations is accidental and not due to a physical link between the events. In Paper~I 
we showed that this is highly unlikely. We did this by demonstrating that the cumulative 
distribution function (CDF) of $|\tau_{\rm corr}|$ can be produced accidentally only with very low 
probability ($\sim 5 \times 10^{-5}$). In other words, if we randomly ``throw rotations'' on the 
{\em Fermi} gamma-ray light curves, it is unlikely that we will get as short time lags as we 
observed.

Here we conduct a similar analysis using the full 3 seasons set of EVPA rotations. We used the 
gamma-ray photon flux curves and the flare fits from Sec.~\ref{subsec:obs_fits}. For  each of the 
rotators with defined $\tau_{\rm corr}$ we randomly selected a JD from the uniform distribution 
within the time range corresponding to the {\em RoboPol} observing season for this blazar. Then we 
identified the closest gamma-ray flare to this JD and obtained $\tau_{\rm corr,sim}$ as was done 
for the observational data. For blazars where two EVPA rotations were observed during a single 
season, we independently performed this procedure twice.
Repeating the experiment $10^6$ times we constructed the CDF of $|\tau_{\rm corr,sim}|$ for each 
trial. In Fig.~\ref{fig:CDF_delays} we show $10^4$ of these CDFs (grey lines) together with 
the observed CDF of $|\tau_{\rm corr}|$ (black). Out of the $10^6$ simulated CDFs only 70 are 
located in their entirety closer to zero than the observed one (i.e., located to the left of the 
observed CDF, shown by the black solid line in Fig.~\ref{fig:CDF_delays}). This result implies that 
the probability of {\em all} time lags in the sample being accidentally so close to zero, as 
observed, is $\sim 7\times10^{-5}$. Thereby, we confirm the results of Paper~I, and we conclude that 
the small time lags we observe suggest a physical link between EVPA rotations and gamma-ray flares.

\section{Correlations between parameters of EVPA rotations and gamma-ray flares} 
\label{sec:correlations}

The results from the analysis in section \ref{sec:rot_gamma} suggest the hypothesis that some (if 
not all) EVPA rotations must be physically related with the nearest gamma-ray flares. If this is 
the case, one would expect that at least some properties of these events are correlated. In the 
following subsections we discuss the results found in the course of this analysis.

\subsection{Flare luminosity vs. rotation amplitude}

We quantify the amplitude of the gamma-ray flare nearest to an EVPA rotation by the measure $L_p$, 
the gamma-ray luminosity at its peak. The amplitude of the rotation, $\Delta \theta_{\rm max}$, is 
simply the difference between the maximum and minimum values of EVPA during the rotation.
\begin{figure}
 \centering
 \includegraphics[width=0.42\textwidth]{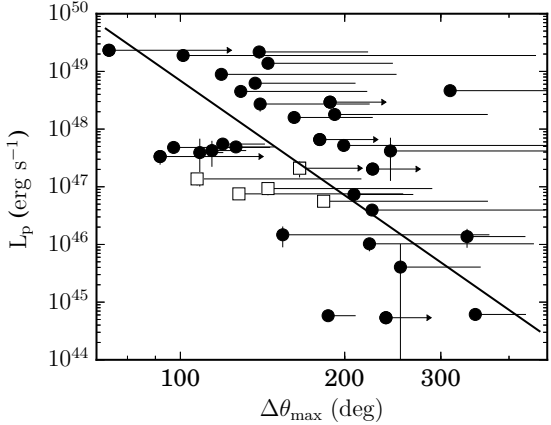}
\caption{Gamma-ray flare luminosity vs. rotation amplitude. The linear fit is shown by the line. 
Empty squares correspond to blazars with uncertain $z$.}
 \label{fig:rot_ampl_vs_gamma_lumin}
\end{figure}
The dependence of $L_p$ on $\Delta \theta_{\rm max}$ is plotted in 
Fig.~\ref{fig:rot_ampl_vs_gamma_lumin}. Hereafter we use logarithmic scales because the functional 
dependence of variables is unknown and it allows to test the general case of power-law dependence, 
which includes also linear correlation. The two quantities in Fig.~\ref{fig:rot_ampl_vs_gamma_lumin} 
are anticorrelated with $r=-0.54$ ($p\text{-value} = 7\times10^{-4}$). The best ordinary least 
squares bisector \citep[OLSB,][]{Isobe1990} fit to the data gives the slope value $-6.6 \pm 1.5$, 
which implies high significance of the correlation.

It should be noted that $\Delta \theta_{\rm max}$ have relatively large uncertainties and in several 
cases they are defined only as lower limits. These uncertainties are caused by observational 
restraints, when either start and/or end of a rotation cannot be pinpointed accurately due to 
insufficient cadence of observations. There is no bias in the values of these uncertainties with 
respect to blazar properties. For instance, they are not correlated to $\Delta \theta_{\rm max}$ or 
$L_p$. Therefore, we ignore these uncertainties in further analysis for simplicity and omit in 
figures for better readability.

Any correlation of $L_p$ with some other parameter may be a manifestation of one of the following 
situations: the parameter under consideration may be correlated with the relative flare amplitude, 
$F_{\rm p}/F_{\rm 3FGL}$, the redshift, $z$, or the beaming properties of the sources i.e., the 
Doppler factor, $\delta$.
We have not found any significant correlation between $\Delta \theta_{\rm max}$ and
$F_{\rm p}/F_{\rm 3FGL}$ ($r=-0.06$, $p\text{-value}=0.72$). However, we indeed found a correlation 
between $\delta$, $z$ and $\Delta \theta_{\rm max}$, as we discuss below.

\subsection{Redshift vs. rotation amplitude}

\begin{figure}
 \centering
 \includegraphics[width=0.42\textwidth]{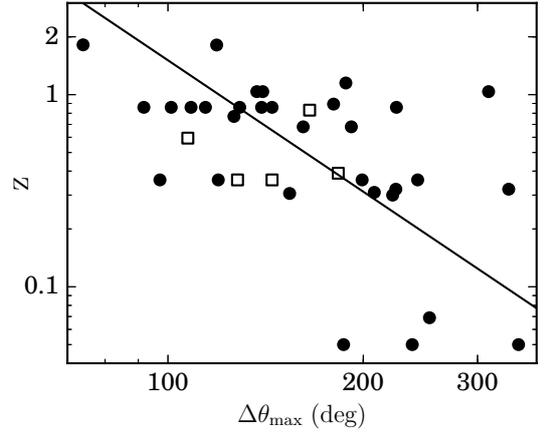}
\caption{Redshift vs. rotation amplitude. The linear fit is shown by the line. Empty squares 
correspond to blazars with uncertain $z$.}
 \label{fig:rot_ampl_vs_z}
\end{figure}
In Fig.~\ref{fig:rot_ampl_vs_z} we show dependence of the redshift on $\Delta \theta_{\rm max}$. The 
two quantities appear to be anticorrelated ($r=-0.56$ and $p\text{-value}=0.001$ for spectroscopic 
$z$), while the OLSB best fit line has a highly significant slope $-2.3\pm0.2$. The correlation 
between $z$ and $\Delta \theta_{\rm max}$ can be caused by a physical relation between $L_p$ and 
$\Delta \theta_{\rm max}$, and the fact that our observing sample suffers from the Malmquist bias 
(i.e. correlation between $z$ and $L_p$), since it is a flux limited sample. Alternatively, $z$ and 
$\Delta \theta_{\rm max}$ can be correlated due to cosmic evolution in the properties of rotations 
(which would induce the correlation of $L_p$ and $\Delta \theta_{\rm max}$, since we tend to see 
more luminous sources at higher redshifts).

\subsection{Jet parameters vs. rotation amplitude}

\begin{figure}
 \centering
 \includegraphics[width=0.42\textwidth]{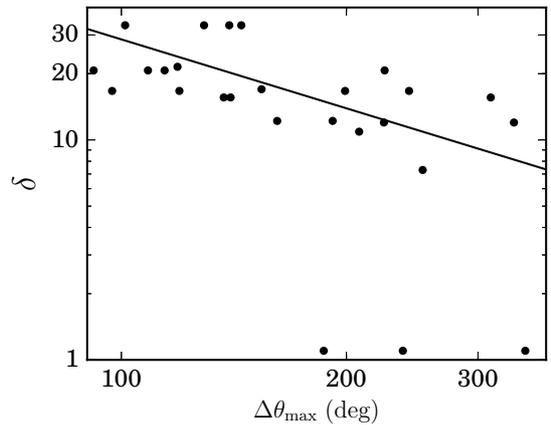}
\caption{Doppler factor, $\delta$, of a blazar vs. rotation amplitude, $\Delta\theta_{\rm max}$. 
The line is a linear fit in logarithmic scale with the three points of 3C~371 excluded.}
 \label{fig:Doppl_vs_rot_ampl}
\end{figure}

Throughout this paper, we use the Doppler factors, $\delta$, derived by \cite{Hovatta2009} from the 
variability of the total flux density at 37 GHz (see Table A1 in Paper~III). In 
Fig.~\ref{fig:Doppl_vs_rot_ampl} we show $\delta$ of rotators, as a function of $\Delta \theta_{\rm 
max}$. 
It shows a  clear anticorrelation of these parameters. This is a surprising result, given the fact 
that the $\delta$  were obtained for an observing period prior to {\em RoboPol} observations under 
the assumption of energy equipartition between the magnetic field and the radiating particles 
\citep{Readhead1994,Lahteenmaki1999}. This assumption of equipartition may not hold in all sources
\citep[see e.g.,][]{Gomez2015,Bruni2017}. The Doppler factors have on average 30\% random errors as 
shown by \cite{Liodakis2015}. Moreover, $\delta$ for the optical emission region may significantly 
differ from $\delta$ of the region emitting at centimetre wavelengths. The three points in 
Fig.~\ref{fig:Doppl_vs_rot_ampl} at $\delta=1.1$, which is approximately order of magnitude lower 
than others, are rotations in 3C~371 (RBPL\,J1806+6949). For this source, the Doppler factor 
in \cite{Hovatta2009} is listed as acceptable, which is their lowest quality class, and the 
sampling of the source was not very good, which could have resulted in an underestimated Doppler 
factor \citep[see ][ for the effects of sampling on the Doppler factor estimates]{Liodakis2015}. 
However, \cite{Fan2013} and \cite{Ghisellini1993} give $\delta=1.36$ and 0.7 consistent with 
\cite{Hovatta2009}. According to \cite{Pesce2001}, this source is an intermediate object 
between BL Lacs and radio galaxies, which supports the suggestion that the source is less beamed.

When 3C~371 is excluded from the analysis, the correlation coefficient between 
$\log(\delta)$ and $\log(\Delta \theta_{\rm max})$ is $r=-0.57$ ($p\text{-value} = 0.005$), while 
the slope of the ordinary least squares bisector regression \citep{Isobe1990} fit is $-1.04
\pm 0.03)$, implying high significance of the correlation.

Additionally, when the gamma-ray luminosity at the flare peak, discussed in the previous 
subsection, is deboosted as $L_{\rm p,jet}=  L_{\rm p}/\delta^{4}$ \citep{Celotti2008} the 
correlation in Fig.~\ref{fig:rot_ampl_vs_gamma_lumin} disappears. The correlation coefficient 
between $L_{\rm p,jet}$ and $\Delta \theta_{\rm max}$ is 0.16 ($p\text{-value} = 0.4$).

The Doppler factor depends on the jet viewing angle, $\alpha$, and the bulk Lorentz factor, 
$\Gamma$, as $\delta = [\Gamma (1-\beta \cos \alpha)]^{-1}$. Therefore, the correlation between 
$\delta$ and $\Delta \theta_{\rm max}$ can be caused by correlation of the latter with 
$\alpha$ and/or $\Gamma$. We examined both possibilities using the estimates of $\alpha$ and 
$\Gamma$ from \cite{Hovatta2009}, where they were calculated from $\delta$ and apparent jet speeds 
derived from Very Long Baseline Interferometry observations.
\begin{figure}
 \centering
 \includegraphics[width=0.42\textwidth]{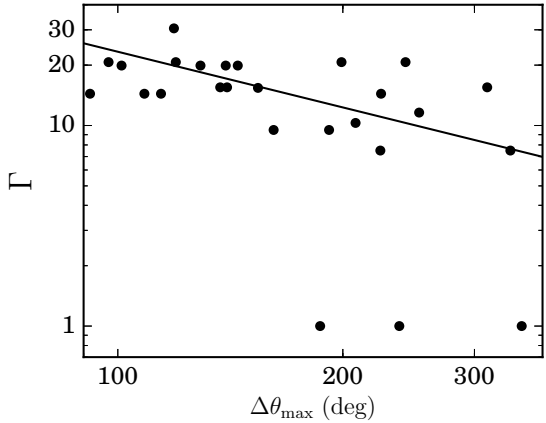}
\caption{Lorentz factor, $\Gamma$, vs. rotation amplitude, $\Delta\theta_{\rm max}$. The line is a 
linear fit in logarithmic scale with the three points at $\Gamma=1$ excluded.}
 \label{fig:Gamma_vs_ampl}
\end{figure}
\begin{figure}
 \centering
 \includegraphics[width=0.42\textwidth]{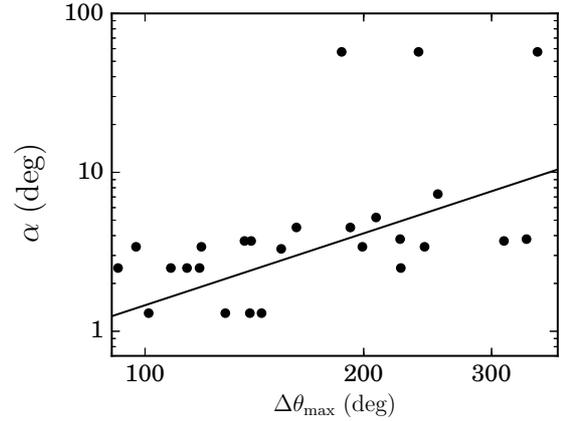}
\caption{Jet viewing angle, $\alpha$, vs. rotation amplitude, $\Delta\theta_{\rm max}$. The line is 
a linear fit in logarithmic scale with the three points at $\alpha=57.3\dg$ excluded.}
 \label{fig:VA_vs_ampl}
\end{figure}
The dependencies of $\Gamma$ and $\alpha$ on $\Delta \theta_{\rm max}$ are shown in 
Fig.~\ref{fig:Gamma_vs_ampl} and \ref{fig:VA_vs_ampl}. Both quantities are marginally correlated 
with $\Delta \theta_{\rm max}$. The orthogonal distance regression fit of $\Gamma$ versus $\Delta 
\theta_{\rm max}$ gives the slope $-0.92\pm0.08$, while the correlation coefficient is $-0.51$ 
($p\text{-value} = 0.01$). Similarly the slope in Fig.~\ref{fig:VA_vs_ampl} is $1.5\pm0.3$, $r=0.51$ 
($p\text{-value} = 0.01$). The outlying points for 3C~371 were excluded from the analysis in both 
cases.

\subsection{Correlation of timescales}

\begin{figure}
 \centering
 \includegraphics[width=0.42\textwidth]{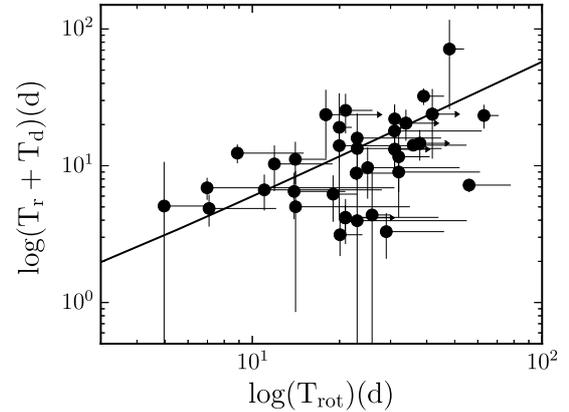}
\caption{Connection of timescales of flares and rotations.}
 \label{fig:Trot_vs_Tflare}
\end{figure}

In addition to amplitudes, we also find that the durations of the gamma-ray flares and EVPA 
rotations are correlated. The dependence of the characteristic timescale of gamma-ray flares, $T_r 
+ T_d$, on the duration, $T_{\rm rot}$ of EVPA rotations is shown in Fig.~\ref{fig:Trot_vs_Tflare}.
The two quantities are positively correlated with $r=0.46$ ($p\text{-value} = 0.005$). The slope of 
the best least squares fit line $0.57\pm0.19$ implies a correlation significance at the $2.8
\sigma$ level. Therefore, we conclude that there is a marginal correlation between the timescales of 
the detected EVPA rotations and the closest gamma-ray flares. There is a link between the EVPA 
rotations and gamma-ray flares. However, there are many more gamma-ray flares in the {\em Fermi} 
light curves in Fig.~\ref{fig:rotations1} -- \ref{fig:rotations3}, so one may ask whether the 
flares associated with the EVPA rotations stand out from the rest in any way.

\section{Are the flares associated with rotations peculiar?}

Visual inspection of the gamma-ray light curves does not reveal any obvious peculiarity of the 
flares closest to EVPA rotations. In most of the cases they appear to be similar in amplitude, 
duration and shape to the rest of the flares occurred during the {\em RoboPol} observing season. To 
investigate whether the flares related to EVPA rotations are not segregated in their parameters 
from other flares, we divided all gamma-ray flares, that were identified and fitted in 
Sec.~\ref{subsec:obs_fits}, into two groups with and without associated EVPA rotation events.

\begin{figure}
 \centering
 \includegraphics[width=0.40\textwidth]{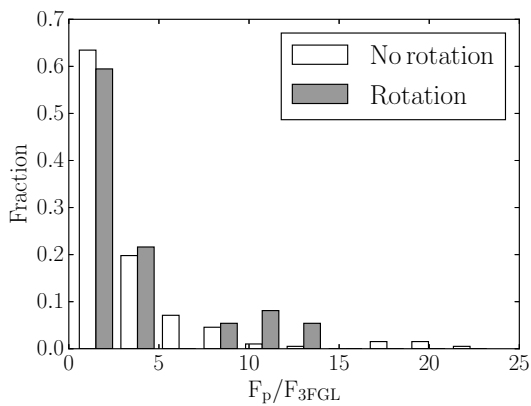}
\caption{Histogram of relative flare amplitudes for the flares closest to EVPA rotations and all 
other flares.}
 \label{fig:hist_rel_ampl}
\end{figure}

In Fig.~\ref{fig:hist_rel_ampl} we show the distribution of relative amplitudes, $F_p/F_{\rm 
3FGL}$, of the flares for the two groups. The K-S test does not reject the null hypotheses that 
both samples are drawn from the same parent population ($p\text{-value} = 0.86$).

\begin{figure}
 \centering
 \includegraphics[width=0.40\textwidth]{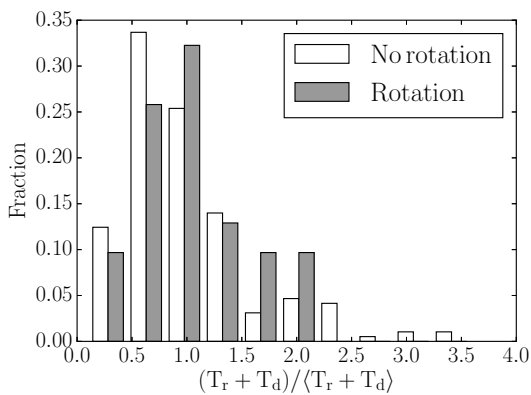}
\caption{Histogram of normalized timescales for the flares closest to EVPA rotations and all other 
flares.}
 \label{fig:hist_rel_dur}
\end{figure}

In order to compare the characteristic timescales of flares in the two groups, we kept only blazars 
with three or more flares (19 out of 24) and found the average timescale, $\langle 
T_r + T_d \rangle$, for each source. Then we normalized $T_r + T_d$ of each flare by this mean 
value. The histogram of this quantity for both groups is shown in Fig.~\ref{fig:hist_rel_dur}. 
According to the K-S test the two distributions cannot be distinguished ($p\text{-value} = 0.26$).

We conclude that gamma-ray flares related to EVPA rotations do not show any peculiar properties. 
Both amplitudes and durations of the flares accompanied by rotations are statistically similar to 
those in flares, where no rotation was detected in {\em RoboPol} data. We emphasize, however, that 
this analysis is done under the assumption that we detected all EVPA rotations that occurred during 
our observations. In fact, observational constraints limit the validity of this assumption. The 
detection efficiency depends on the rotation rate and on the cadence of observations (see Sec.~3.1 
of Paper III for details).


\section{Discussion and conclusions} \label{sec:conclusion}

In this paper we investigated the connection between optical EVPA rotations and gamma-ray activity 
in blazars. We used the data from the {\em RoboPol} programme i.e., the full 3-year monitoring and 
the 40 detected rotations. This is the largest number of rotations detected in a uniform way for a 
statistically complete sample of blazars so far. The study was possible not only due to {\em 
RoboPol} data, but also thanks to the availability of continuous gamma-ray data obtained by {\em 
Fermi}-LAT.

In Paper~I, based on the analysis of the correlation between the EVPA rotations and gamma-ray flares 
detected during the first observing season, we had found an indication of a bi-modal distribution 
of $\tau_{\rm corr}$: large amplitude gamma-ray flares appeared to be closely linked with 
EVPA rotations, while smaller amplitude flares were not. This dependence was considered as a 
manifestation of two types of rotations coexisting in blazars. In this work we demonstrated that 
this difference is caused by the less accurate determination of the flare peak position in the case 
of flares with low gamma-ray photon statistics.

The major result of our study is that the EVPA rotations are physically linked with gamma-ray 
flares. We measured the average delay of $\overline{\tau_{\rm corr}}=1.5\pm3.1$ d, which is fully 
consistent with zero. The uncertainty of $\overline{\tau_{\rm corr}}$ is mainly determined
by the uncertainty in the determination of $t_p$ of the low amplitude gamma-ray flares. This 
uncertainty will not improve significantly in the near future, but the uncertainty of the time lag 
could be reduced by an increase in the number of detected EVPA rotations in blazars. We found that 
the probability that this small time lag between gamma-ray flares and EVPA rotations is accidental 
is $\sim 7\times10^{-5}$.

This result and the significant correlations we detected between the parameters of the flares and 
rotations (Sec.~\ref{sec:correlations}), confirm the physical connection between these events. 
Indeed, we found that both the ``amplitude'' and ``duration'' of these events are correlated. The 
significant anticorrelation between the luminosity of the gamma-ray flares and the amplitude of the 
rotations $\Delta \theta_{\rm max}$ is explained by stronger relativistic boosting in blazars that 
exhibited lower amplitude EVPA rotations i.e., when $\delta$ is larger (flare luminosity is larger) 
$\Delta \theta_{\rm max}$ is smaller. This dependence, in turn, is explained by the dependence of 
$\Delta \theta_{\rm max}$ on the bulk Lorentz factor and the viewing angle of the jet. The faster 
the jet and the smaller the viewing angle, the lower the amplitude of the rotation. The 
characteristic timescale of the gamma-ray flares shows a marginal ($\sim 2 \sigma$) positive 
correlation with the duration of corresponding EVPA rotations: longer gamma-ray flares seem to be 
associated with longer rotations.

Our results strongly favour the deterministic nature of EVPA rotations. If a substantial fraction 
of these events was produced by a random walk of the polarization vector, then both amplitudes and 
timescales of rotations would be random quantities independent from corresponding properties of 
gamma-ray flares. In this case the correlations in Figs.~\ref{fig:rot_ampl_vs_gamma_lumin} - 
\ref{fig:Trot_vs_Tflare} would be smeared out. The connection between parameters of EVPA rotations 
and gamma-ray flares also implies that these events are produced in the same region of the jet. 
Therefore, EVPA rotations can be used for localization of the gamma-ray emission zone within the 
jet. However, the possibility that there is an underlying stochastic process producing a random 
walk of the polarization vector, superposed with a deterministic process that produces the large 
rotations contemporaneously with gamma-ray flares, cannot be ruled out based on our data 
\citep{Kiehlmann2017}. In fact, along with the measurement uncertainties, the presence of random 
walk events could be responsible for the rather wide spread of points in 
Fig.~\ref{fig:rot_ampl_vs_gamma_lumin} - \ref{fig:Trot_vs_Tflare}.

Finally, we showed that the gamma-ray flares related to EVPA rotations do not show any distinctive 
properties in their amplitudes or characteristic timescales when compared to other flares that 
occurred during {\em RoboPol} observations. In principle, every gamma-ray flare could be 
accompanied by an EVPA rotation, and the absence of a recorded rotation in {\em RoboPol} data could 
be due to sparse sampling. In order to investigate whether every gamma-ray flare is indeed 
accompanied by a swing in EVPA, a continuous, very high cadence monitoring of known rotators is 
required.

%
%
%
%
%
%
%

\section*{Acknowledgements}

The {\em RoboPol} project is a collaboration between Caltech in the USA, MPIfR in Germany, 
Toru\'{n} Centre for Astronomy in Poland, the University of Crete/FORTH in Greece, and IUCAA in 
India. The U. of Crete group acknowledges support by the ``RoboPol'' project, which is implemented 
under the ``Aristeia'' Action of the  ``Operational Programme Education and Lifelong Learning'' and 
is co-funded by the European Social Fund (ESF) and Greek National Resources, and by the European
Comission Seventh Framework Programme (FP7) through grants PCIG10-GA-2011-304001 ``JetPop'' and
PIRSES-GA-2012-31578 ``EuroCal''. This research was supported in part by NASA grant NNX11A043G and 
NSF grant AST-1109911, and by the Polish National Science Centre, grant number 2011/01/B/ST9/04618.
S.\,K. is supported by NASA grant NNX13AQ89G. T.\,J.\,P. and A.\,C.\,S.\,R. acknowledge support 
from NASA award NNX16AR41G. K.\,T. and G.\,V.\,P. acknowledge support by the European Commission 
Seventh Framework Programme (FP7) through the Marie Curie Career Integration Grant 
PCIG-GA-2011-293531 ``SFOnset''. M.\,B. acknowledges support from NASA Headquarters under the 
NASA Earth and Space Science Fellowship Program, grant NNX14AQ07H. We acknowledge the hard work by 
the \textit{Fermi}-LAT Collaboration that provided the community with unprecedented quality data and 
made Fermi Science Tools so readily available. We thank members of the \textit{Fermi}-LAT 
Collaboration for their useful comments that improved the manuscript.





\bibliographystyle{mnras}
\bibliography{bibliography_manual}

\bsp 
\label{lastpage}
\end{document}